\documentclass[11pt,USenglish]{article}
\usepackage[utf8]{inputenc}
\pdfoutput=1

\usepackage{jheppub}
\usepackage{setspace}

\usepackage[USenglish]{babel}
\usepackage{verbatim}
\usepackage{prettyref}
\usepackage{amsmath}
\usepackage[normalem]{ulem}
\usepackage{amssymb}
\usepackage{graphicx}
\usepackage{setspace}
\usepackage{esint}
\usepackage{empheq}
\usepackage{physics}

\usepackage{tikz}
\newcommand*{\boxcolor}{black}
\makeatletter
\renewcommand{\boxed}[1]{\textcolor{\boxcolor}{%
\tikz[baseline={([yshift=-1ex]current bounding box.center)}] \node [rectangle, minimum width=1ex,rounded corners,draw] {\normalcolor\m@th$\displaystyle#1$};}}
 \makeatother

\makeatletter
\newcommand{\vast}{\bBigg@{3}}
\makeatother

\title{Six-point functions and collisions in the black hole interior}

\author{Felix M.\ Haehl,}
\author{Alexandre Streicher,}
\author{Ying Zhao}
\affiliation{School of Natural Sciences, Institute for Advanced Study,\\
1 Einstein Drive, Princeton, NJ 08540, USA}
\emailAdd{haehl@ias.edu}
\emailAdd{streicher@ias.edu}
\emailAdd{zhaoying@ias.edu}

\abstract{In the eternal AdS black hole geometry, we consider two signals sent from the boundaries into the black hole interior shared between the two asymptotic regions. We compute three different out-of-time-order six-point functions to quantify various properties of the collision of these signals behind the horizons: $(i)$ We diagnose the strength of the collision by probing the two-signal state on a late time slice with boundary operators. $(ii)$ We quantify two-sided operator growth, which provides a dual description of the signals meeting in the black hole interior, in terms of the quantum butterfly effect and quantum circuits. $(iii)$ We consider an explicit coupling between the left and right CFTs to make the wormhole traversable and extract information about the collision product from behind the horizon. At a technical level, our results rely on the method of eikonal resummation to obtain the relevant gravitational contributions to Lorentzian six-point functions at all orders in the $G_N$-expansion. We observe that such correlation functions display an intriguing factorization property. We corroborate these results with geodesic computations of six-point functions in two- and three-dimensional gravity.}

\begin{document}

\maketitle

\section{Introduction}

The use of quantum information theoretic concepts in studies of the AdS/CFT duality has proven to be an extraordinarily fruitful approach towards quantum gravity. One of the most basic, yet most consequential lessons has been the realization that entanglement between CFT degrees of freedom manifests itself in terms of a gravitational description \cite{Ryu:2006bv}. In particular, the maximally entangled `thermofield double' state, which purifies the thermal density matrix of a given CFT, has been argued to be dual to an eternal black hole geometry with an interior that is shared between the system and its purification \cite{Maldacena:2001kr,VanRaamsdonk:2010pw,Maldacena:2013xja}. The shared interior geometry presents a puzzling aspect of this `ER=EPR' paradigm, as it allows for signals sent from the decoupled boundary systems to meet even though there are no boundary interactions \cite{Marolf:2012xe}.
In this paper we discuss how this seemingly mysterious meeting is encoded in the dual CFT. We study the meeting in the interior by probing the entangled structure of the state with various six-point functions. 
We will consider the eternal AdS black hole geometry with two signals sent into the wormhole, one from each boundary, and consider several setups to probe the resulting state.

\subsection{Overview and summary of results}

The theme of this paper is to use thermal six-point functions of pairwise identical operators to probe properties of the wormhole, which provides a gravitational description dual to the thermofield double state \cite{Maldacena:2001kr,VanRaamsdonk:2010pw,Hartman:2013qma},
\begin{equation}
\label{eq:TFDdef}
|\text{TFD}\rangle = \sum_k e^{-\frac{\beta}{2} \, E_k} \, |E_k \rangle_L \otimes |E_k \rangle_R \;\; \in \; {\cal H}_{\text{CFT}_L} \otimes {\cal H}_{\text{CFT}_R} \,.
\end{equation}

\begin{figure}
\begin{center}
\includegraphics[width=0.9\textwidth]{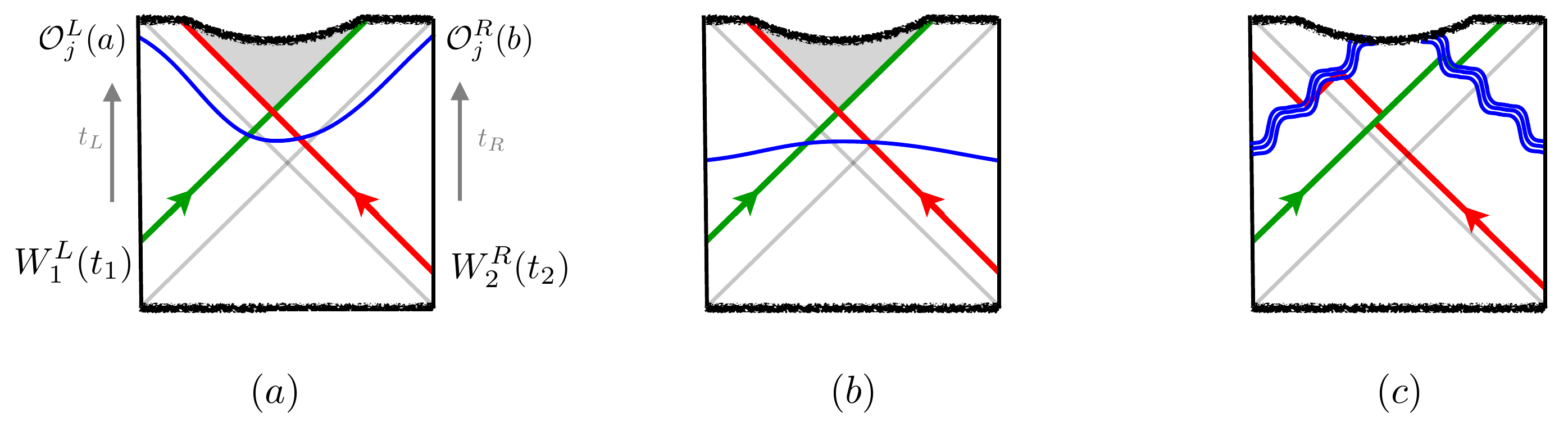}
\end{center}
\vspace{-.3cm}
\caption{The three setups we consider in this paper. $(a)$ Scattering experiment in the eternal AdS black hole geometry dual to the thermofield double state. Two perturbations are inserted, one on the left and one on the right, and we probe the state using a two-point function of ${\cal O}_j$ on a later time slice. $(b)$ The same setup but with ${\cal O}_j$ operators inserted at $a=b=0$ serves to quantify the two-sided scrambling process and the `overlap' of the two growing perturbations. $(c)$ A coupling between the left and right theories modifies the time evolution in such a way that the wormhole becomes traversable and we can extract information about the scattering process behind the horizon.} 
\label{fig:bulkSetup}
\end{figure}

The basic idea is that the highly entangled structure of this state provides a pattern of correlations between the two (non-interacting) CFTs, which can be probed using correlation functions. In gravity correlation functions of highly boosted operators are well approximated by scattering events involving shockwave geometries \cite{DRAY1985173,Sfetsos:1994xa,Shenker:2013pqa,Shenker:2013yza}. 
We will discuss three different six-point function configurations, which carry information about different aspects of the wormhole geometry dual to \eqref{eq:TFDdef}. Let us summarize these at a qualitative level (see figure \ref{fig:bulkSetup} for illustrations):
\begin{itemize}
    \item \textbf{(a) Setting up a collision and diagnosing it:} Our first setup concerns sending two signals into the wormhole -- one from the left and one from the right. Each signal corresponds to two operator insertions (one each for the bra and the ket state describing the perturbed geometry). The correlations between left and right boundary systems are highly sensitive to the collision behind the horizon. In order to diagnose whether or not the signals met inside the wormhole before falling into the singularity, we therefore compute a probe two-point function between the two boundaries at a late time.
    \item \textbf{(b) Characterizing the overlap of the perturbations:}
    A single particle falling towards the horizon has been interpreted in terms of the growth of its dual operator under unitary time evolution due to the quantum butterfly effect \cite{Susskind:2018tei}.\footnote{ The notion of operator size growth that we employ can be made concrete in the Sachdev-Ye-Kitaev (SYK) model \cite{Kitaev:2015aa,Maldacena:2016hyu}, where a concrete decomposition of generic operators into basis operators exists \cite{Qi:2018bje} and is closely related to the gravitational description in AdS${}_2$ \cite{Lin:2019qwu}.} It was also argued that the trajectory of the infalling particle in the interior close to the horizon is a manifestation of the growth of the operator in the quantum circuit stored in the interior \cite{Hartman:2013qma,Susskind:2014moa,Zhao:2020gxq,Haehl:2021sib}. In this paper, we will show that two particles meeting in the interior can be interpreted in terms of the overlap of largely independent growth of two operators in a single shared circuit: the two growing operators correspond to perturbations `infecting' the qubits of the circuit in opposite directions \cite{Zhao:2020gxq,HaehlZhao:essay}. The operator size of the thermofield double with two perturbations thus is related to the number of quantum gates unaffected by both.
    \item \textbf{(c) Extracting the collision product from the wormhole:} Finally, we may ask if it is possible to extract information about the result of the collision from the wormhole. In other words: can we make the wormhole traversable such that one of the signals reaches the opposite boundary {\it after} colliding with the other signal? Again, each signal corresponds to two operator insertions. To make the wormhole traversable, one in addition needs to perform time evolution with an operator pair that couples the left and right boundary systems \cite{Gao:2016bin,Maldacena:2017axo,Maldacena:2018lmt}.\footnote{ Strictly speaking, we will consider the exponential of an operator pair coupling the two boundaries. But since we can treat the many scattering events as independent, we will reduce the problem to  six-point functions again.} For a suitable choice of left-right coupling, the creation of negative energy density deforms the geometry in a way that opens the wormhole for transport. However, this negative energy needs to be carefully balanced against the positive energy created by the additional perturbation.
\end{itemize}
All three of our investigations involve Lorentzian six-point functions of the schematic form 
\begin{equation}
  {\cal F}_6 \sim \frac{\langle  W_1 W_1 \, {\cal O}_j {\cal O}_j \, W_2 W_2 \rangle }{\langle W_1 W_1 \rangle \langle {\cal O}_j {\cal O}_j \rangle \langle W_2 W_2  \rangle }
\end{equation}
for different time orderings and different insertion times (all expectation values refer to the thermofield double state). The operators $W_{1,2}$ will be associated with (potentially strong) perturbations sent into the geometry. The operator ${\cal O}_j$ will either play the role of a probe diagnosing properties of the signals and their collision (setups $(a)$ and $(b)$), or as a piece of the `teleportation operator' (setup $(c)$).

Let us note an interesting structure, which we will observe in many cases: our results for six-point functions take the form 
\begin{equation}
\label{eq:F6factorize}
   {\cal F}_6
   \sim \frac{\langle  W_1 W_1 \, {\cal O}_j {\cal O}_j  \rangle }{\langle W_1 W_1 \rangle \langle {\cal O}_j {\cal O}_j\rangle }
   \times\frac{\langle   {\cal O}_j {\cal O}_j \, W_2 W_2 \rangle }{\langle {\cal O}_j {\cal O}_j \rangle \langle W_2 W_2  \rangle } 
   \times\frac{\langle  W_1 W_1   W_2 W_2 \rangle }{\langle W_1 W_1 \rangle  \langle W_2 W_2  \rangle } \times {\cal F}_{6,\text{conn.}} 
\end{equation}
where the four-point factors inherit their time orderings from those of ${\cal F}_6$. The {\it `connected' factor} will be our particular object of interest, as it encodes important physical effects in all three examples discussed above.\footnote{ Let us elaborate on the subscript ``conn". While we will not prove it in generality, we expect the following picture to be true: the subscript refers to a diagrammatic expansion of the six-point function in terms of Schwarzian mode (or graviton) exchange diagrams. The contribution ${\cal F}_{6,\text{conn.}}$ originates from connected diagrams. To order $G_N^2$ this agrees with the usual notion of large $N$ factorization (see also Appendix \ref{sec:perturbative}). At higher orders in $G_N$ the usual notion of large $N$ factorization is expected to apply {\it before} exponentiation, i.e., for the quantity $\log {\cal F}_6$.}  However, its precise time dependence and the way in which it competes (or doesn't compete) with the `four-point factors' will be different depending on the setup:
\begin{itemize}
    \item In the first setup (diagnosing the collision) all the four-point factors are ${\cal O}(1)$ and play no important role. All the time-dependent information about the collision is encoded in ${\cal F}_{6,\text{conn.}}$: if the operators sourcing the left and right signals are inserted too late, no collision occurs and ${\cal F}_{6,\text{conn.}}\approx 1$, indicating that the full six-point diagnostic ${\cal F}_6$ is just a product of independent propagators. On the other hand, for signals sent early, there is a strong collision and ${\cal F}_{6,\text{conn.}} \approx 0$, indicating a large deviation from the factorized result. For intermediate times, the connected piece interpolates in a way that quantifies details about the collision strength.
    \item In the second setup (characterizing the overlap of the perturbations) the four-point factors $\langle W_1W_1{\cal O}_j {\cal O}_j\rangle$ and $\langle {\cal O}_j {\cal O}_j W_2W_2\rangle$ are non-trivial and characterize the scrambling (or operator size growth) of each signal separately. The connected six-point piece stays order $1$. This approximate factorization of the six-point function into a product of two four-point functions reflects the fact that that the two perturbations grow independently from each other in a shared quantum circuit.\footnote{ We will comment later on the small corrections to this picture due to mild but nontrivial time-dependence of ${\cal F}_{6,\text{conn.}}$, which is interesting from the point of view of six-point scrambling.}
    \item In the third setup (making the wormhole traversable) the full correlator is more complicated, and all three four-point factors individually have strong time dependence. While factorization for two of the four-point factors still occurs naturally, the question whether or not the full structure \eqref{eq:F6factorize} persists is obscured by the more complicated setup. See appendix \ref{app:traversable} for more details.
\end{itemize}

\subsection{The structure of gravitational interactions}

Thermal six-point functions have been studied in recent years for several reasons: $(i)$ In suitable out-of-time-order correlation functions (OTOCs), they serve as fine-grained diagnostics of the quantum butterfly effect \cite{Haehl:2017pak}, generalizing the well known four-point measures of chaos \cite{Larkin:1969aa,Shenker:2013pqa,Roberts:2014ifa,Kitaev:2015aa}. $(ii)$ Six-point functions of a combination of heavy and light operators encode entropic measures of quantum entanglement \cite{Banerjee:2016qca}, diagnose  chaos in the context of the eigenstate thermalization hypothesis \cite{Anous:2019yku}, etc. $(iii)$ The perturbative gravitational contributions to six-point functions provides a detailed testing ground for uncovering the structure of gravitational scattering processes and identity conformal blocks \cite{Haehl:2019eae,Anous:2020vtw}. For other recent work on six-point functions and conformal blocks in the context of AdS/CFT and thermal systems, see \cite{Chen:2016dfb,Anous:2017tza,Rosenhaus:2018zqn,Jepsen:2019svc,Jensen:2019cmr,Kusuki:2019gjs,Alkalaev:2020kxz,Fortin:2020ncr,Hoback:2020pgj,Fortin:2020yjz,Kundu:2021bub}.

With few exceptions in simple cases, all of these works on six-point functions have one drawback in common: the gravitational contributions to the correlation functions are treated perturbatively.\footnote{ In the language of conformal blocks this usually means that, either one focuses on global conformal blocks, or on leading contributions at large central charge.} Physically, this results in a restriction of the regime of validity of various computations. For example, in the context of six-point OTOCs, perturbative contributions display exponential growth, but they are not sufficient to observe thermalization (saturation due to higher order effects) after a scrambling time. A technical advancement in the present paper is that we will overcome this issue. We will use the eikonal formalism \cite{HOOFT198761,Verlinde:1991iu,Kabat:1992tb,Kiem:1995iy,Cornalba:2007zb,Brower:2007qh} in the form developed recently in \cite{Shenker:2014cwa,Maldacena:2017axo} to resum the leading exponentially growing contributions to the scattering process described by the six-point function. Such contributions have an exponential time dependence, which is balanced by powers of Newton's constant, i.e., $(G s)^n$, where $s$ stands for a suitable center of mass energy. Perturbatively, such contributions originate from graviton `ladder diagrams'. We will also take the opportunity to discuss the relation between the eikonal resummation technique and the perturbative approach, which can be made quite precise in the simplest context of two-dimensional Jackiw-Teitelboim gravity \cite{TEITELBOIM198341,JACKIW1985343} and its  description in terms of the Schwarzian action \cite{Almheiri:2014cka,Kitaev:2015aa,Maldacena:2016upp}. We argue for the intriguing structure of \eqref{eq:F6factorize}, hidden in the gravitational contributions to the six-point function, which becomes evident in the out-of-time-order configuration: the six-point function factorizes into a product of four-point contributions and `connected' six-point terms.

It is tempting to interpret the structure of \eqref{eq:F6factorize} as a generalized notion of large $N$ factorization of gravitational contributions to Lorentzian correlation functions. The reason it emerges can be traced back to the {\it eikonalization} of gravitational scattering amplitudes, which takes into account effects at arbitrary orders in the perturbative gravitational expansion, as long as they are compensated for by large relative boost factors. The eikonal method that we will use to calculate six-point functions in the various cases is designed to extract such contributions.

A similar structure was observed for Virasoro identity blocks in two-dimensional conformal field theories in \cite{Anous:2019yku,Anous:2020vtw}. In that case the four-point factors would be Virasoro four-point identity blocks and ${\cal F}_{6,\text{conn.}}$ would correspond to connected `diagrams' in the effective theory of  stress tensor exchanges. It is interesting to speculate about a general eikonalization argument of gravitational interactions (or Virasoro identity blocks) based on the perturbative structure of reparametrization mode exchanges. We elaborate on this idea in appendix \ref{sec:perturbative}, but leave a detailed investigation for the future.

\subsubsection*{Outline}

This paper is organized as follows. In section \ref{sec:collision} we describe the collision setup in detail and discuss general properties of the six-point function used to probe the state resulting from the collision. We also confirm our results from the bulk perspective, using geodesic approximations in JT gravity and AdS${}_3$ gravity. In section \ref{sec:strength} we consider the second setup, which serves to quantify the amount by which the perturbations `overlap' in the quantum circuit picture. We also offer an interpretation in terms of operator size (in the case where the boundary dynamics can be simulated by the SYK model).
Then, we turn to the traversable wormhole setup in section \ref{sec:traversable}, arguing that one can extract information about the collision from the wormhole. In section \ref{sec:calculations} we discuss the eikonal resummation calculations underlying most of these results, both from the point of view of momentum space and in position space. We also give results for some six-point functions with general insertion times, commenting on their general properties and their features in gravity. We summarize open questions in section \ref{sec:conclusion} and delegate technical details to several appendices.

\section{Diagnosing collisions in the wormhole interior}
\label{sec:collision}

Our first setup concerns sending signals into both black holes and diagnosing whether a collision happens in the interior. 
We discuss a particular six-point function appropriate for this task, and compute it using eikonal methods as well as gravitational calculations in asymptotically AdS${}_2$ and AdS${}_3$ geometries. The setup in this section is largely based on \cite{HaehlZhao:essay} and here we will provide more details and discussions.

\subsection{Setup and general properties}
\label{sec:setup}

We would like to diagnose features of the collision of two perturbations sent into the wormhole dual to the state \eqref{eq:TFDdef} from the left and right boundaries, respectively. The perturbed state of interest is therefore of the form
\begin{equation}
\label{eq:stateDef}
  W_2^R(t_2) W_1^L(t_1) |\text{TFD}\rangle \,,
\end{equation}
where $W_{1,2}$ are two CFT operators with dimensions $\Delta_{1,2}$, whose superscripts refer to whether they are inserted in the left or right CFT. For simplicity we will assume $W_{1,2}^\dagger = W_{1,2}$. We suppress explicit labels indicating the spatial insertion points (often we only discuss one-dimensional boundary theories). The setup is illustrated in figure \ref{fig:bulkSetup}.

Having understood the left-right perturbed state, we need to probe it. The simplest probe we can consider is a two-point function of some operator ${\cal O}_j$ between the two boundaries. Such a two-point function will be sensitive to correlations between the left and right systems and it will develop an interesting time dependence in the presence of the perturbations described above. The two-sided two-point function of ${\cal O}_j$ is\footnote{ See also \cite{Shenker:2014cwa} for a similar setup.}
\begin{equation}
\label{eq:FtildeDef}
{\cal F}_6(t_1,t_2;a,b) = {\cal N} \, \Big( \langle \text{TFD} | W_1^{L}(t_1) W_2^R(t_2) \Big) {\cal O}_j^L(a) {\cal O}_j^R(b) \Big(   W_2^R(t_2) W_1^L(t_1) |\text{TFD}\rangle \Big)
\end{equation}
where ${\cal N}$ is a normalization factor specified below (see, e.g., \eqref{eq:F6def}). Note also that -- if possible -- it can be more convenient to compute not just a single two-point function, but sum over all two-point functions of a suitable set of basis operators. This will be particularly natural in systems with a finite number of degrees of freedom such as the SYK model (see section \ref{sec:strength}).

The detailed time dependence of the six-point correlation function \eqref{eq:FtildeDef} will illuminate the physics associated with the meeting of the two perturbations in the interior of the wormhole. Intuitively, we expect that the behavior of ${\cal F}_6$ should be different for $t_{1,2} \gtrsim 0$ compared to $0> t_{1,2}$: in the former case the two perturbations do not meet in the interior before hitting the singularity, while in the latter case they do collide and there exists a post-collision region (shaded grey in figure \ref{fig:bulkSetup}).

\begin{figure}
\begin{center}
\includegraphics[width=\textwidth]{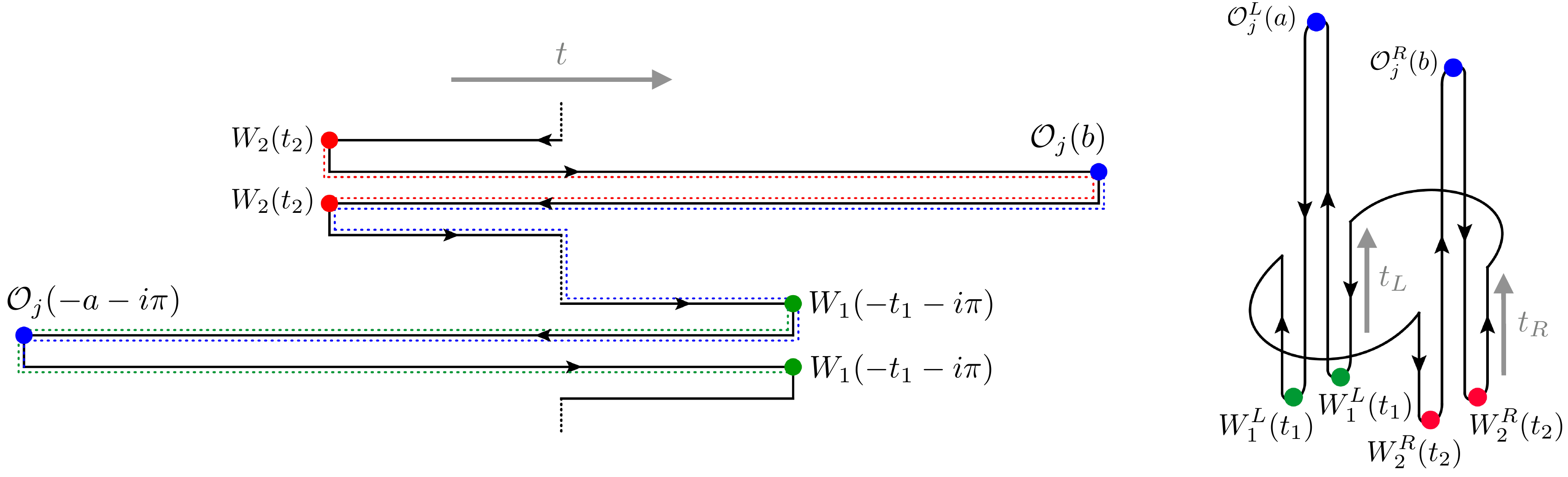}
\end{center}
\vspace{-.3cm}
\caption{Illustration of the `irreducibly' out-of-time-order configuration \eqref{eq:F6def}. We show the complex time contours where the three perturbations have support, as well as the left-right configuration, which is inspired by the bulk picture. The complex contour time coordinate is $t_R$ and $-t_L - i\pi$ for right and left boundary times, respectively.} 
\label{fig:contour6fold}
\end{figure}

In the following it will be useful to write ${\cal F}_6$ more explicitly as a correlation functions involving the thermal density matrix $\rho$. Also introducing a natural normalization, the central object of interest in this paper will thus be the following Lorentzian correlation function:\footnote{ Note that in this one-sided picture the left time has the opposite orientation of the right time. When translating between left-right correlators (such as \eqref{eq:FtildeDef}) and analytically continued one-sided correlators (such as \eqref{eq:F6def}), we need to flip the sign of insertion times on the left.}
\begin{equation}
\label{eq:F6def}
\boxed{\;\;
  {\cal F}_6(t_1,t_2;a,b) \equiv \frac{\text{tr} \left\{ W_1(-t_1) {\cal O}_j(-a) W_1 (-t_1) \, \rho^{\frac{1}{2}}\, W_2(t_2){\cal O}_j(b)  W_2(t_2) \, \rho^{\frac{1}{2}} \right\}_\delta}{ \text{tr} \left\{ W_1W_1\, \rho \right\}_\delta  \;\text{tr} \left\{ {\cal O}_j(-a) \rho^{\frac{1}{2}}{\cal O}_j(b)\, \rho^{\frac{1}{2}} \right\}_\delta \; \text{tr} \left\{ W_2W_2\, \rho \right\}_\delta}
  \;\;}
\end{equation}
 The subscripts $\delta$ on the expectation values indicates that the insertions are UV regulated by a point splitting procedure: neighboring operators are separated by an additional small imaginary time.
We will take large $a,b$ in this section. The arrangement of operators in this correlation function is illustrated in figure \ref{fig:contour6fold}.

\subsection{Result for the late time six-point function}

Let us now give the results for ${\cal F}_6$ in the configuration \eqref{eq:F6def}. The derivation of these results will be provided later, in section \ref{sec:calculations}.

\subsubsection*{Result for heavy operators}

The result takes the simplest form when $W_{1,2}$ are much heavier than ${\cal O}_j$. In this case, we can treat the particles created by $W_{1,2}$ as probes and find the following result (see \eqref{eq:F6calc2appendix}):
\begin{equation}
\label{eq:F6saddle1}
\boxed{\;\;
{\cal F}_6(t_1,t_2;a,b) \approx  \left[ \frac{1}{1+ \frac{G^2\Delta_1 \Delta_2}{16 \sin \delta_1 \sin \delta_2} \, e^{-(t_1+t_2)}} \right]^{2\Delta_j} \qquad (a,b \gg -t_{1,2}) 
\;}
\end{equation}
where we set $\beta = 2\pi$, as we will often do. First note that for large $a,b$, this quantity only depends on $t_1+t_2$. 
The exponential time dependence of this quantity is purely a six-point effect and cannot be attributed to disconnected four-point contributions. We can already see some of the crucial features of ${\cal F}_6$ in this configuration: it only decays to $0$ for large {\it negative} $t_1+t_2$, but approaches an asymptotic value of $1$ for {\it positive} $t_1+t_2$. The former case corresponds to very early pearturbations, which lead to an high energy collision near the bifurcation surface (see figure \ref{fig:bulkSetup}). The exponential decay for negative $t_1+t_2$ happens over a characteristic timescale 
\begin{equation}
  2 t_* = \frac{\beta}{2\pi} \log \left[ \frac{16\sin \delta_1 \sin \delta_2}{G^2\Delta_1 \Delta_2} \right] \,,
\end{equation}
which is {\it twice} the scrambling time. This is, of course, related to the fact that we are computing a six-point function sensitive to more fine-grained aspects of quantum chaos \cite{Haehl:2017pak}.

\subsubsection*{Geodesic approximation in gravity}

In gravity, we can compute the approximate six-point function using a geodesic limit.
As explained above, ${\cal F}_6$ can be thought of as a left-right two-point function in a state that consists of two perturbations acting on the thermofield double background. Such a two-point function can be approximated in terms of the geodesic distance $d$ between the boundary insertion points of ${\cal O}_j$:
\begin{align}
\label{eq:F6geodesic1}
	{\cal F}_6(t_1,t_2;a,b) \approx {\cal F}_6^\text{(geodesic)}(t_1,t_2;a,b) \equiv \frac{e^{-\Delta_j\, d_\text{pert.} } }{e^{-\Delta_j\,  d_\text{unpert.}} }\,,
\end{align}
where we normalized the geodesic distance in the perturbed geometry, $d_\text{pert.}$, by the geodesic distance in the unperturbed background. 
We expect this approximation to be good when ${\cal O}_j$ can be considered as a probe in the background generated by $W_{1,2}$. This corresponds to the assumption that $\Delta_j \ll \Delta_{1,2}$. In this case, the perturbed state corresponds to a geometry with two shockwaves (similar to figure \ref{fig:bulkSetup}). 
In appendices \ref{app:JTderivation} and \ref{app:3D geometry} we compute the geodesic distances in the perturbed and unperturbed geometries in JT gravity and in AdS${}_3$ gravity.  We quote here the final expressions when $a,b$ are large.\vspace{2mm}

\noindent\underline{JT gravity result:} In two-dimensional JT gravity, we find (see \eqref{eq:JTfinalResult}):
\begin{equation}
  {\cal F}_6^\text{(geodesic)} \approx \qty[\frac{1}{1+\frac{\delta S_1\delta S_2}{4(S-S_0)^2}\,e^{-(t_1+t_2)}}]^{2\Delta_j} \qquad (a,b \gg -t_{1,2})\,,
  \label{collision_diagnosis}
\end{equation}
where $S-S_0$ is the above-extremal black hole entropy, and the entropy differentials due to the shockwaves are related to the operator dimensions and regulators as follows:
\begin{equation}
\label{eq:SDelta}
\frac{\delta S_1}{S-S_0} =\frac{G\Delta_1}{2\sin\delta_1} = \frac{\Delta_1}{C \sin \delta_1}\,,\qquad \frac{\delta S_2}{S-S_0} =\frac{G\Delta_2}{2\sin\delta_2}= \frac{\Delta_2}{C \sin \delta_2} \,.
\end{equation}
Note that \eqref{collision_diagnosis} is precisely the same as the eikonal result in the saddle point approximation, \eqref{eq:F6saddle1}.\vspace{2mm}

\noindent\underline{AdS${}_3$ gravity result:}
The probe geodesic approximation \eqref{eq:F6geodesic1} applies also in three-dimensions as the assumption $1 \ll \Delta_j \ll \Delta_{1,2}$ continues to suppress fluctuations.
For simplicity we work with zero-momentum operators, which will approximately source symmetric, null shocks in the BTZ geometry. Then, pasting vacuum spacetimes together along null surfaces solves Einstein's equations while supporting the non-zero stress-energy along said shocks.\footnote{ The only complication is avoiding conical singularities at the intersection of multiple shocks. Thus, there must be a trivial holonomy around any such intersection, which constrains the possible geometries.} We explain this calculation in detail in appendix \ref{app:3D geometry}.
For the general result with arbitrary $a,b$, see \eqref{eq:AdS3FullResult}. 
For $a,b\gg -t_{1,2}$ the result takes exactly the same form as above. Indeed, we find precisely the expression \eqref{collision_diagnosis} with $S_0 = 0$. 

\subsubsection*{Full eikonal result for large $a,b$}

We now give a more detailed result for the eikonal integral \eqref{eq:F6intAbstract}. The only assumption we make is $a,b \gg -t_{1,2}$ such that the left-right probe correlator \eqref{eq:BsDef02} simplifies.\footnote{ See also \eqref{eq:BsDef}.} In particular, we do not assume a hierarchy of operator dimensions anymore. In appendix \ref{app:eikonal1} we derive the following result:
\begin{equation}
\label{eq:F6calc5}
\begin{split}
{\cal F}_6 
&\approx 
 \frac{ \Gamma(2\Delta_{1j})\Gamma(2\Delta_{2j})}{\Gamma(2\Delta_1)\Gamma(2\Delta_2)} \; \frac{1}{z^{4\Delta_j}} \;  {}_1F_2\left( 2\Delta_j ; \, 1- 2 \Delta_{1j},\,1-2\Delta_{2j} ;\,- \frac{1}{z^2} \right) \\
&\;\; +  \frac{\Gamma(2\Delta_{j1})\Gamma(2\Delta_{21})}{\Gamma(2\Delta_j)\Gamma(2\Delta_2)} \,  \frac{1}{z^{4\Delta_1}} \; {}_1F_2\left( 2\Delta_1 ; \, 1- 2 \Delta_{j1},\,1-2\Delta_{21} ;\, - \frac{1}{z^2}  \right) \\
  &\;\; +   \frac{\Gamma(2\Delta_{j2})\Gamma(2\Delta_{12})}{\Gamma(2\Delta_j)\Gamma(2\Delta_1)} \, 
  \frac{1}{z^{4\Delta_2}} \; {}_1F_2\left( 2\Delta_2 ; \, 1- 2 \Delta_{j2},\,1-2\Delta_{12} ;\, - \frac{1}{z^2} \right) 
\end{split}
\end{equation}
for $a,b \gg -t_{1,2}$, where we abbreviated $\Delta_{ab} = \Delta_a - \Delta_b$ and we defined
\begin{equation}
z^2 = \frac{G^2}{64 \sin\delta_1\sin\delta_2} \,e^{-(t_1+t_2)}\,.
\end{equation}
\begin{figure}
\begin{center}
\includegraphics[width=0.6\textwidth]{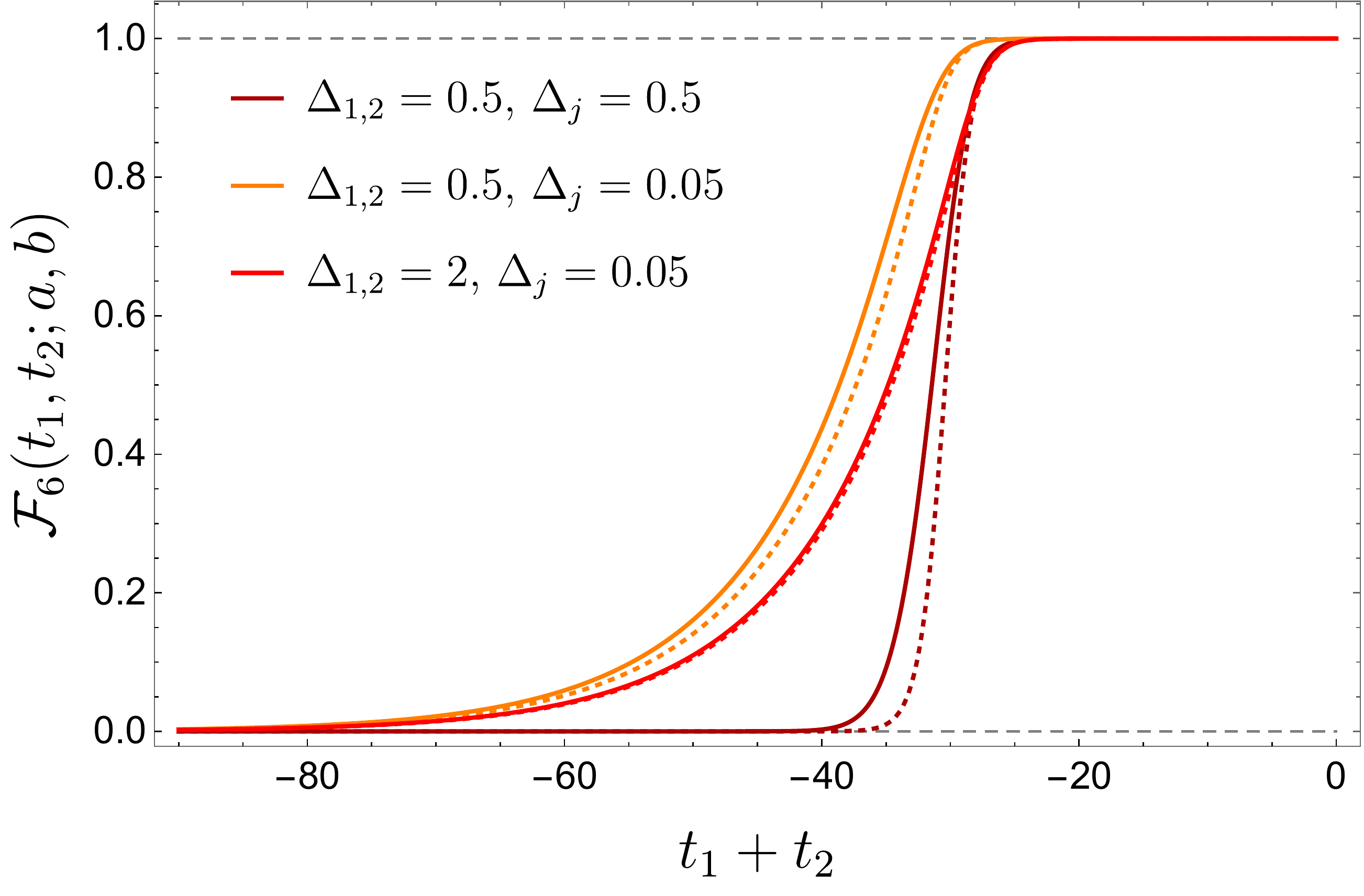}
\end{center}\vspace{-.5cm}
\caption{Plot of ${\cal F}_6(t_1,t_2;a,b)$ for $a,b \gg -t_{1,2}$. Full lines give the exact result \eqref{eq:F6calc5} for the eikonal integral. Dotted lines correspond to the `heavy-heavy-light' saddle point approximation \eqref{eq:F6saddle1}, which also coincides with the geodesic approximation in JT gravity. The saddle point approximation gets better as the ratio $\Delta_{1,2} / \Delta_j$ increases. The characteristic timescale for exponential decay is $-(t_1+t_2) \sim 2 t_*$. We set $G/(2 \sin \delta_{1,2}) = 10^{-6}$ (such that $2t_* \approx 28$).}
\label{fig:F6plot1}
\end{figure}
Note that \eqref{eq:F6calc5} is symmetric under permutations of $\Delta_i$. Further note that despite appearances, \eqref{eq:F6calc5} has a finite limit when any of the operator dimensions become equal.
The generalized hypergeometric function has the following limiting behavior for early insertion times:\footnote{ Similarly, for $z \ll 1$, one can show:
\begin{equation}
\label{eq:1F2asymp}
  {}_1F_2\left(2\Delta;\, \alpha_1,\alpha_2; \, -\frac{1}{z^2}\right) \sim
  \frac{\Gamma(\alpha_1)\Gamma(\alpha_2)}{\Gamma(\alpha_1-2\Delta)\Gamma(\alpha_2-2\Delta)} \, z^{4\Delta} 
  + \frac{\Gamma(\alpha_1)\Gamma(\alpha_2)}{\sqrt{\pi}\, \Gamma(2\Delta)} \; z^{-\frac{1}{2} - 2\Delta+\alpha_1+\alpha_2} \; \cos \frac{2}{z}
\end{equation}
Note that for $\Delta_1 + \Delta_2 + \Delta_j \geq \frac{3}{2}$ the small-$z$ behavior is not well-behaved. This reflects the fact that the eikonal approximation assumed from the beginning that $-(t_1+t_2) \sim {\cal O}(2t_*)$. Many expressions are valid beyond this regime, but there is no a priori reason they had to be. In practice, this is not an issue because the small-$z$ behavior of ${\cal F}_6$ outside of the regime of validity of the eikonal approximation is known anyway.}
\begin{equation}
\begin{split}
  {}_1F_2\left( 2\Delta;\, \alpha_1,\alpha_2; \, -\frac{1}{z^2}\right) &\sim 1 - \frac{2\Delta}{\alpha_1\alpha_2} \, \frac{1}{z^2} + \ldots \qquad (z \gg 1) 
\end{split}
\end{equation}

In figure \ref{fig:F6plot1} we plot \eqref{eq:F6calc5} for various values of operator dimensions. We also show the saddle point (or geodesic) approximation \eqref{eq:F6saddle1}. The essential feature is again the following: for positive $t_1+t_2$ no collision inside the wormhole occurs and ${\cal F}_6$ takes the value $1$. This corresponds to the factorized answer, where the three pairs of operators do not interact in any detectable way and the six-point function simply factorizes into two-point functions. On the other hand, when $-(t_1+t_2)$ approaches $2t_*$, a strong collision in the wormhole interior happens. The six-point function is sensitive to this event, decorrelates, and eventually approaches 0.

\section{Properties of a quantum circuit with overlapping perturbations}
\label{sec:strength}

Once we can diagnose the happening of the collision, we wish to quantify more detailed properties of the collision event. From the perspective of `ER=EPR' \cite{Maldacena:2001kr,Maldacena:2013xja}, the collision can be described in terms of the overlap of two perturbations in the quantum circuit acting on the EPR pairs (see figure \ref{circuit_5}) \cite{Zhao:2020gxq}. 
The larger the overlap is, the stronger the collision is. The properties of the overlap region in the quantum circuit have previously been related to the geometry of the post-collision region in the bulk dual. The most direct way to study such properties is by means of operator growth in the presence of two perturbations.

\subsection{Review of operator growth and quantum circuit model}

The growth of one single operator in the quantum circuit was well studied \cite{Susskind:2014jwa,Roberts:2014isa,Roberts:2018mnp, Qi:2018bje}.  For concreteness, consider the SYK model with $N$ Majorana fermions $\psi_j$. The size $n_\infty[{\cal O}]$ of some operator ${\cal O}$ built out of the fermions can be defined as the average number of fermion flavors occurring in the operator \cite{Roberts:2018mnp}:
\begin{equation}
  n_\infty[{\cal O}] \equiv \frac{1}{4} \sum_j \text{tr} \left( \{ {\cal O} ,\psi_j \}^\dagger\, \{{\cal O},\psi_j \} \right) \,,
\end{equation}
This notion of operator size can be upgraded to a measure of size in thermal states \cite{Qi:2018bje}:
\begin{equation}
\label{eq:nBetaDef}
 \frac{n_\beta[{\cal O}]}{n_{max}} \equiv \frac{n_\infty[{\cal O}\rho^{\frac{1}{2}}] - n_\infty[\rho^{\frac{1}{2}}] }{n_{max} - n_\infty[\rho^{\frac{1}{2}}]} \,,
\end{equation}
where $n_{max} = \frac{N}{2}$ is the size of a completely scrambled operator. The thermal size \eqref{eq:nBetaDef} of a single fermion, $n_\beta[\psi_1(t)]$, is well known to be related to an out-of-time-ordered four-point function \cite{Qi:2018bje}, which grows exponentially until it saturates to $n_{max}$ at a time of order $t \gtrsim t_*$:
\begin{equation}
    \label{eq:4point}
    \mathcal{F}_4^\text{(SYK)}(t)\equiv 1-\frac{n_{\beta}[\mathcal{O}(t)]}{n_{max}} = - \frac{\sum_j\tr(\mathcal{O}^{\dagger}(t)\psi_j\mathcal{O}(t)\rho^{\frac{1}{2}}\psi_j\rho^{\frac{1}{2}})}{\sum_j\tr(\mathcal{O}^{\dagger}(t)\mathcal{O}(t)\rho)\tr(\psi_j\rho^{\frac{1}{2}}\psi_j\rho^{\frac{1}{2}})}
\end{equation}

\begin{figure}
\begin{center}
\includegraphics[width=0.9\textwidth]{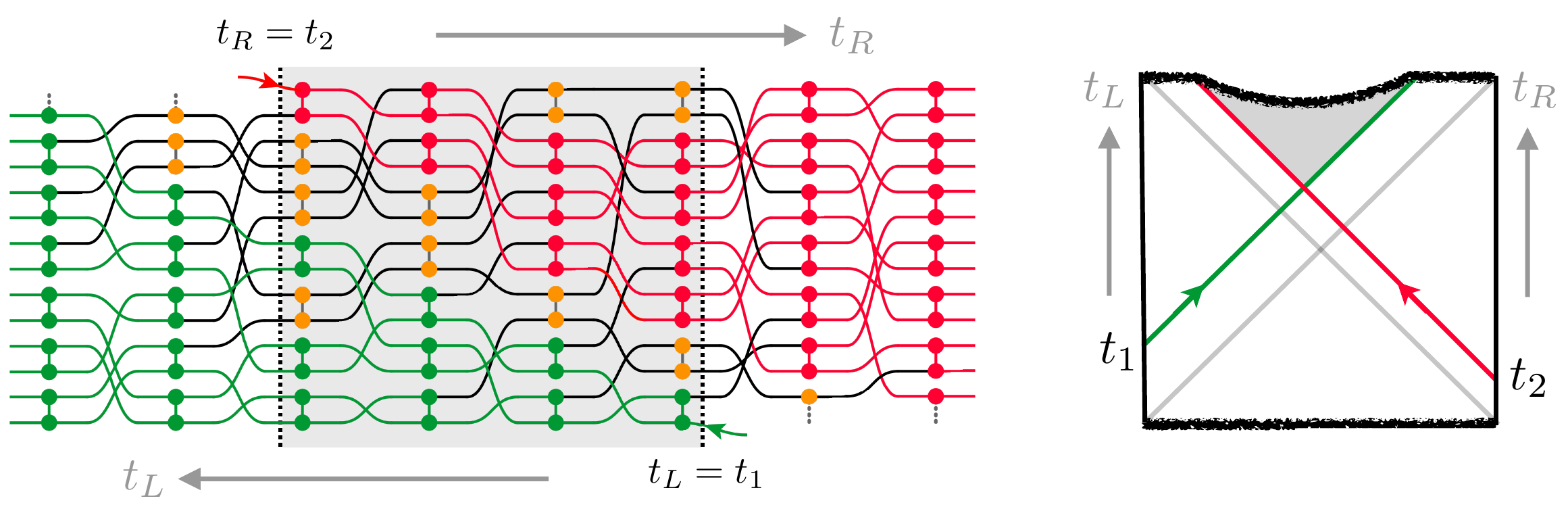}
\end{center}
\vspace{-.3cm}
\caption{Via `ER=EPR' the quantum circuit model provides a qubit picture for the evolution of the thermofield double state. Time evolution on the left (right) boundary systems corresponds to increasing the depth of the circuit towards the left (right). The two signals sent into the wormhole correspond to two perturbations (`infections') spreading in opposite directions of the circuit. We quantify the amount of overlap of the two perturbations in the circuit (shaded grey in the left panel).}
\label{circuit_5}
\end{figure}

The epidemic model of operator growth posits that the size of the operator can be understood in terms of the perturbation spreading through the quantum circuit, thus `infecting' an exponentially increasing number of quantum gates (c.f., figure \ref{circuit_5}) \cite{Susskind:2014jwa}. More precisely the quantity ${\cal F}_4(t)$ corresponds to the percentage of healthy gates at circuit time $t$.

\subsection{Operator growth for two perturbations}
\label{sec:twoPert}

The six-point function \eqref{eq:F6def} can be understood in terms of the thermal size of an operator with two perturbations present. To make this precise, we define an SYK instance ${\cal F}_6^\text{(SYK)}$ of the six-point function as follows: first, we replace our generic `probe' operators by SYK fermions: $W_1 \rightarrow \psi_1$ and $W_2 \rightarrow \psi_2$.\footnote{ As noted before, summing over all fermions is natural in the context of operator size: we want to detect the overlap of the perturbed state with {\it all} elementary fermions weighted equally. See also \cite{HaehlZhao:essay}.} It is then easy to check that we can write the six-point function as follows:
\begin{equation}
\label{eq:6point}
\boxed{\;\;
 {\cal F}_6^\text{(SYK)}(t_1,t_2;0,0) =  1- \frac{n_\infty[  \psi_1(-t_1) \rho^{\frac{1}{2}} \psi_2(t_2)  ] - n_\infty[\rho^{\frac{1}{2}}] }{n_{max} - n_\infty[\rho^{\frac{1}{2}}]}
 \;}
\end{equation}
That is, the six-point function \eqref{eq:F6def} with $a = b = 0$ can be understood in terms of the `size' of the operator 
\begin{equation}
\label{eq:OpDef}
 \psi_1(-t_1) \rho^{\frac{1}{2}} \psi_2(t_2) \end{equation}
relative to the thermal background $\rho^{\frac{1}{2}}$. This combination is easy to understand operationally: it corresponds to perturbing the thermofield double state with two different fermions $\psi_1^L(t_1)$ and $\psi_2^R(t_2)$ from the left and the right.

From the epidemic model of operator growth we expect that, the probability of being healthy at a given circuit time when both perturbations are present is equal to the probability of being healthy from the first epidemic times the probability of being healthy from the second epidemic. This simple observation implies a non-trivial relation between six-point and four-point functions. 
We saw before that the percentage of healthy gates with one perturbation is given by the four-point function \eqref{eq:4point}. Similarly, the percentage of healthy gates in the presence of two perturbations is given by the six-point function \eqref{eq:6point}. Due to the independent spread of the perturbations, we would thus predict:
\begin{align}
\label{eq:approxfac}
    \mathcal{F}_6^{\text{(epidemic)}}(t_1, t_2; 0,0)\approx \mathcal{F}_4(t_1)\mathcal{F}_4(t_2)\,.
\end{align}

\begin{figure}
\begin{center}
\includegraphics[width=0.57\textwidth]{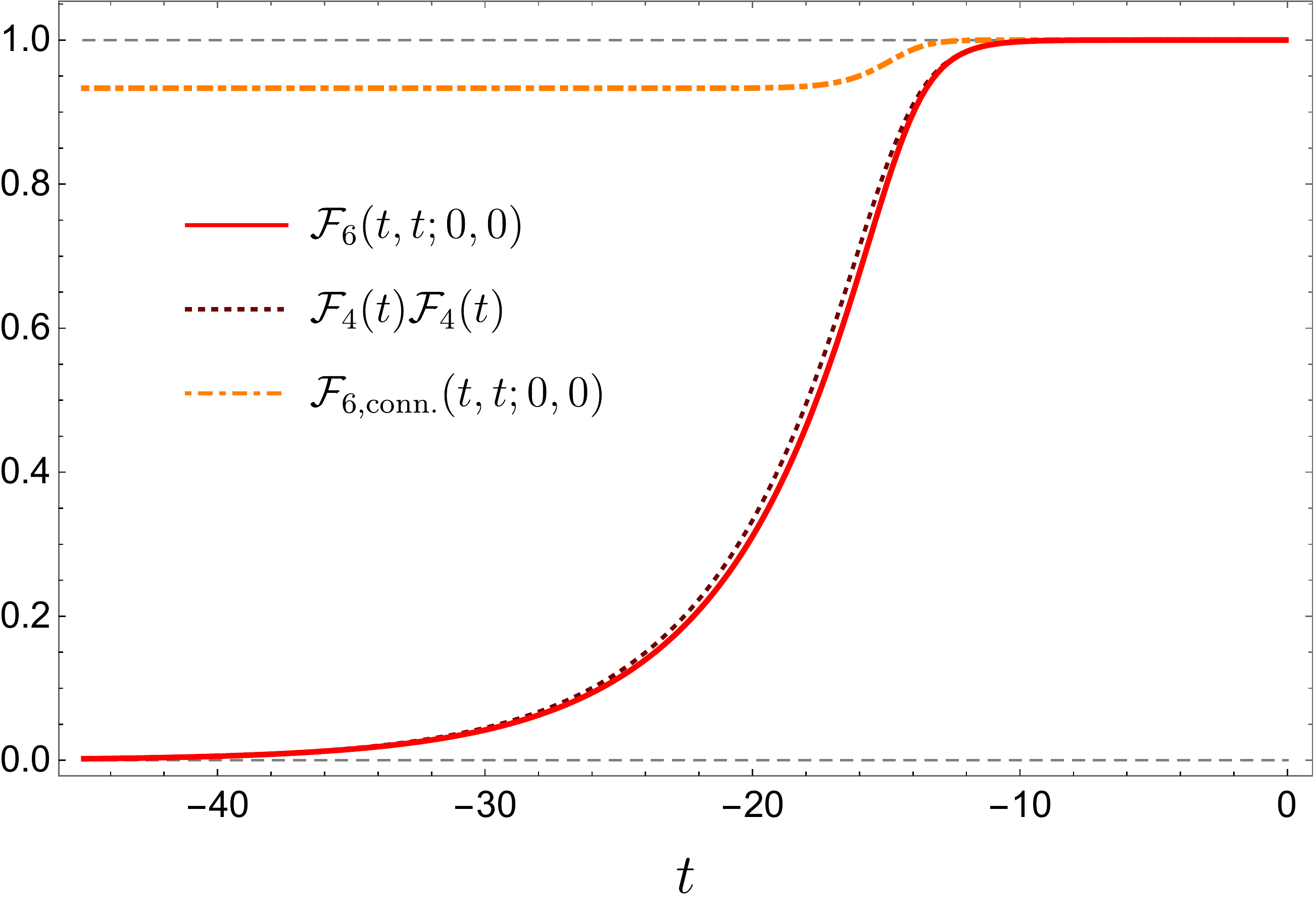}
\end{center}\vspace{-.5cm}
\caption{Illustration of the result \eqref{eq:F6calc2} for $t_1 = t_2 \equiv t$. Because the `connected' contribution saturates to an ${\cal O}(1)$ floor value, it only changes the full six-point function by a very small amount. The transition happens when both $-t_1 \sim t_*$ and $-t_2 \sim t_*$. We set $\Delta_{1,2} = 2$, $\Delta_j=0.05$, and $G/(2\sin\delta_{1,2}) = 10^{-6}$ (such that $t_* \approx 13.8$).}
\label{fig:F6plotEpidemic}
\end{figure}

We now have the tools to study $\mathcal{F}_6(t_1, t_2;0,0)$ in detail using the eikonal method. The contour to compute $\mathcal{F}_6(t_1, t_2; 0,0)$ is given in figure \ref{fig:contour6fold}.  The methods developed to compute ${\cal F}_6(t_1,t_2;a,b)$ in the previous section can easily be adapted to the situation where $a=b=0$. In particular, from the eikonal calculation in the approximation where $\Delta_{1,2} \gg \Delta_j$ (see section \ref{sec:calculations}), we obtain the following result:
\begin{figure}
\begin{center}
\includegraphics[width=0.9\textwidth]{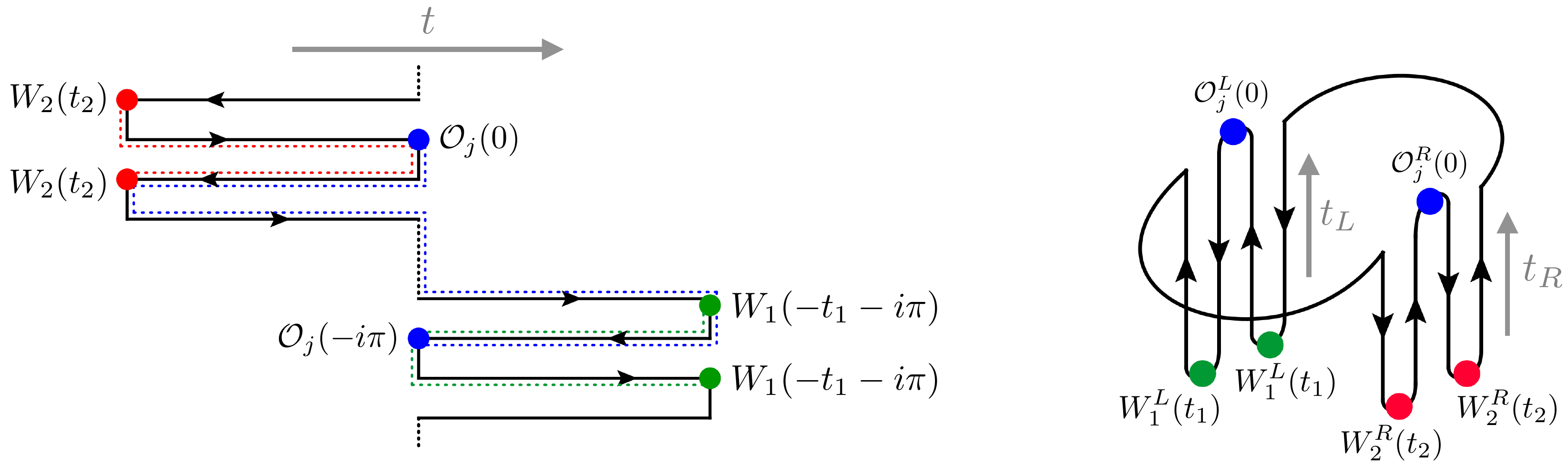}
\end{center}
\vspace{-.3cm}
\caption{Six-point configuration ${\cal F}_6$ corresponding to the case $a=b=0$ (`maximally', but not `irreducibly' out-of-time-order). We show both the contour-time picture, and the left-right configuration.} 
\label{fig:contour6foldSymm}
\end{figure}
{\small
\begin{equation}
\label{eq:F6calc2}
\boxed{\;\;
{\cal F}_6(t_1,t_2;0,0) 
   \approx
\left[ \frac{1}{1+G\Delta_1 \, z_1} \right]^{2\Delta_j} \left[ \frac{1}{1+ G\Delta_2 \, z_2} \right]^{2\Delta_j} 
\left[ \frac{1}{1+ \frac{ G^2\Delta_1\Delta_2 \, z_1 \, z_2 }{ \left( 1+ G\,\Delta_1 \, z_1 \right) \left( 1+ G\,\Delta_2 \, z_2 \right)}} 
\right]^{2\Delta_j} 
\;}
\end{equation}
}\normalsize
where 
\begin{equation}
  z_1 = \frac{e^{-t_1} }{8\, \sin \delta_1 } \,,\qquad
  z_2 = \frac{e^{-t_2} }{8\, \sin \delta_2 }  \,.
\end{equation}
Note that (as advertised in the introduction), \eqref{eq:F6calc2} is of the form
\begin{equation}
\label{eq:F6factorize2}
  {\cal F}_6(t_1,t_2;0,0) = {\cal F}_4(t_1) \, {\cal F}_4(t_2) \, {\cal F}_{6,\text{conn.}}(t_1,t_2;0,0)
\end{equation}
where ${\cal F}_{6,\text{conn.}}$ is the connected contribution that cannot be attributed to four-point effects. It stays order $1$ in the current setup, such that we reproduce the prediction of the epidemic model, \eqref{eq:approxfac}.\footnote{ Due to the qualitative nature of the epidemic model, it does not capture the fact that ${\cal F}_{6,\text{conn.}}$ decays from 1 to $2^{-2\Delta_j}$.} The relation \eqref{eq:approxfac} was used in \cite{Zhao:2020gxq} to estimate the total number of healthy gates in the overlap region and was shown to correspond to the spacetime volume of the post-collision region. Here, we understand at a more detailed level that this relation is a result of the property of the connected piece of the six-point function staying order $1$. We illustrate these features in figure \ref{fig:F6plotEpidemic}.

We also note that the result \eqref{eq:F6calc2} can again be obtained from geodesic calculations in gravity. For JT gravity, the general result \eqref{eq:JTfinalResult} immediately implies \eqref{eq:F6calc2} in the case where $a=b=0$. Similarly, the general geodesic result in AdS$_3$ gravity, \eqref{eq:AdS3FullResult}, implies a similar structure in three dimensions.

\subsection{Comments on six-point scrambling}

The configuration discussed here (figure \ref{fig:contour6foldSymm}) was previously suggested as a six-point probe of quantum chaos \cite{Haehl:2017pak,Haehl:2018izb,Anous:2020vtw}. In those works it was argued that the exponential time dependence of the connected factor ${\cal F}_{6,\text{conn.}}$ serves as a diagnostic of scrambling, which is sensitive to more fine-grained details of the dynamics than the four-point functions. In addition to the initial exponential decay of ${\cal F}_{6,\text{conn.}}$ from its maximum value 1, we now observe a surprising new effect, which only manifests after resumming $(Gz_i)^n$-contributions using the eikonal method: the connected six-point factor never decays to zero, but saturates at a floor value,
\begin{equation}
    {\cal F}_{6,\text{conn.}}(t_1,t_2;0,0) \longrightarrow 2^{-2\Delta_j} \qquad\quad (-t_{1,2} \gg t_*) \,.
\end{equation}
This feature is perhaps unexpected from observing the initial exponential decay, but it is crucial for consistency with notions of operator growth and the predictions of the epidemic model in the previous subsection. This demonstrates a novel feature of six-point scrambling in this out-of-time-order configuration. This observation also suggests that ${\cal F}_6(t_1,t_2;a,b)$ for large $a,b$ might serve as more immediate six-point diagnostic of scrambling: as shown in \eqref{eq:F6saddle1}, this quantity is in fact dominated by its connected contribution. Furthermore it exponentially decays from 1 to 0, thus being more analogous to the properties of the four-point OTOCs.

At a perturbative level, we note that the factorization \eqref{eq:F6factorize2} and the full eikonal result \eqref{eq:F6calc2} (including the connected factor) provide a window into the structure of gravitational interactions. In appendix \ref{sec:perturbative} we discuss in some detail how the structure \eqref{eq:F6calc2} can be understood from a perturbative treatment of the Schwarzian theory of reparametrizations. In particular we show that a triple expansion of \eqref{eq:F6calc2} in the parameters $G\Delta_1$, $G\Delta_2$ and $\Delta_j$ can be understood order by order as perturbative Schwarzian mode exchanges. Terms originating from the `four-point' factors ${\cal F}_4$ are associated with `disconnected' Schwarzian mode exchange diagrams. Only `connected' diagrams contribute to ${\cal F}_{6,\text{conn.}}$. The eikonal result \eqref{eq:F6calc2} therefore provides a detailed check of the structure \eqref{eq:F6factorize}.

\section{Traversable wormholes: extracting information about the collision}
\label{sec:traversable}

Having understood the physical significance of the collision and ways to detect it using the six-point function ${\cal F}_6$, we can ask an even more ambitious question: is it possible to not only {\it detect} the collision from outside, but to {\it extract} the collision product from the wormhole? The mechanism by which this is possible was first explored in \cite{Gao:2016bin}. Our discussion follows more closely the methods developed in \cite{Maldacena:2017axo}.

\subsection{Traversable wormhole setup with collision}

\begin{figure}
\begin{center}
\includegraphics[width=0.4\textwidth]{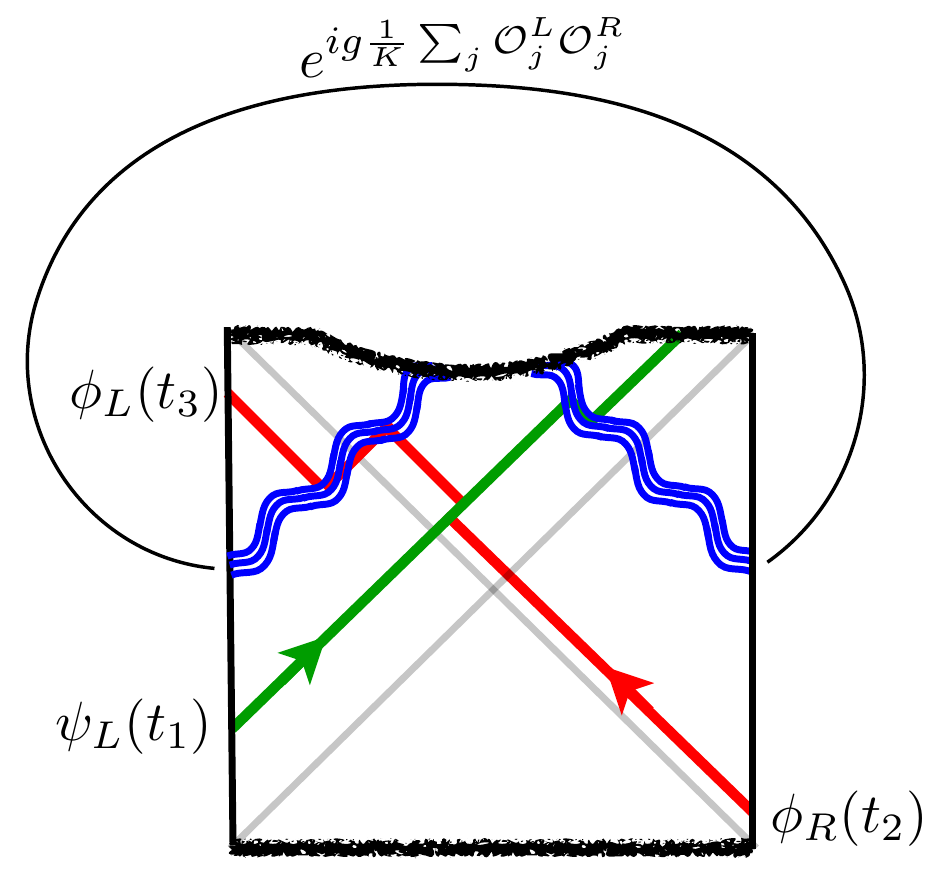}
\end{center}
\vspace{-.3cm}
\caption{Traversable wormhole setup: the `signal' $\phi$ collides with the `perturbation' $\psi$ and subsequently exits the wormhole due to a negative energy shock created by the left-right coupling $\exp [ig \frac{1}{K}\sum_{j = 1}^K{\cal O}_j^L {\cal O}_j^R]$. This way, we can extract information about the collision from behind the horizon.}
\label{traversable_Penrose}
\end{figure}

Starting from thermofield double, we apply the perturbation $e^{i\epsilon_L\psi^L}$ at time $t_1$ on the left. The signal sent from the right is modelled by applying the operator $e^{i\epsilon_R\phi^R}$ at time $t_2$. With appropriately chosen $t_1$ and $t_2$, the two signals will meet in the interior and collide. In order to enable the signal $\phi$ to traverse the wormhole after the collision, and reach the left boundary (with information about the collision product), we apply the `teleportation operator' 
\begin{equation}
e^{igV} \equiv e^{ig\frac{1}{K}\sum_{j = 1}^K{\cal O}^L_j(t){\cal O}^R_j(t)}\,,    
\end{equation} 
which couples the two sides at time $t$ (without loss of generality we will set $t=0$). To detect whether the protocol succeeded, and to measure the collision product at the left boundary, we compute the expectation value of $\phi^L$ on the left side at a later time $t_3$. This one-point function in the presence of the perturbation and the teleportation operator takes the following form:
\begin{align}
	&\expval{e^{-i\epsilon_L\psi^L}e^{-i\epsilon_R\phi^R}e^{-igV}\phi^L e^{igV}e^{i\epsilon_R\phi^R}e^{i\epsilon_L\psi^L}}\\
	=\ &\expval{e^{-igV}\phi^L e^{igV}}-i\epsilon_R\expval{\qty[\phi^R, e^{-igV}\phi^L e^{igV}]}+i\epsilon_R\epsilon_L^2\expval{\qty[\phi^R, \psi^L e^{-igV}\phi^L e^{igV}\psi^L]}+\ldots \nonumber
\end{align}
Our goal is to detect a signal that depends {\it both} on $\epsilon_R$ (meaning the signal has traversed the wormhole), as well as $\epsilon_L$ (meaning the signal has encountered a collision in the interior). The third term in the above expansion is the first nontrivial indicator of both these processes. We will therefore compute the commutator
\begin{align}
    \expval{[\phi^R, \psi^L e^{-igV}\phi^Le^{igV}\psi^L]} \equiv -2i \, \langle \phi^R \phi^L \rangle \langle \psi^L\psi^L \rangle \, \text{Im}\,\mathcal{C} \,,
\end{align}
where the normalized thermal correlator is
\begin{align}
    \mathcal{C} = \frac{\bra{\text{TFD}}\psi^L e^{-igV}\phi^Le^{igV}\psi^L\phi^R\ket{\text{TFD}}}{\bra{\text{TFD}} \phi^R \phi^L \ket{\text{TFD}}\bra{\text{TFD}}\psi^L\psi^L \ket{\text{TFD}}}\,.
\end{align}
Another way to think about this commutator is as the {\it response} $\phi^L$ to a signal $\phi^R$, where the signal is time evolved subject to both the teleportation operator and the effects of the perturbation $\psi^L$. The competition between these two evolution operators will determine whether or not the signal exits the wormhole and gives a nontrivial response.

To compute ${\cal C}$, we first make the following approximation:
\begin{align}
\label{eq:TFDpsiApprox}
    \bra{\text{TFD}}\psi^Le^{-igV}\approx \bra{\text{TFD}}\psi^Le^{-ig\expval{V}_{\psi}}
\end{align}
where $\langle V \rangle_\psi$ is the expectation value of $V$ in the presence of the perturbation $\psi_L$,\footnote{ This is, of course, just the well-known expression for a conformal two-point function in a `heavy' state.}
\begin{equation}
    \expval{V}_{\psi} = \left[ \frac{1}{2+ \frac{G\Delta_\psi }{4 \sin\delta}\,e^{-t_1} } \right]^{2\Delta_j} \,.
\end{equation}
This approximation is good when the perturbation $\psi_L$ is not too early ($|t_1|\ll t_*$).\footnote{Later we will see that in order for the protocol to succeed, the perturbation $\psi_L$ cannot be too early and the collision cannot be strong.}
Using this approximation, we have
\begin{align}
\label{eq:CCtilde0}
    \mathcal{C} \approx \frac{e^{-ig\expval{V}_{\psi}}}{\langle \phi^R\phi^L \rangle\langle \psi^L\psi^L \rangle}\, \big{\langle}\psi^L(t_1)\phi^L(t_3)e^{igV}\phi^R(t_2)\psi^L(t_1)\big{\rangle}\,.
\end{align}
\begin{figure}
\begin{center}
\includegraphics[width=0.8\textwidth]{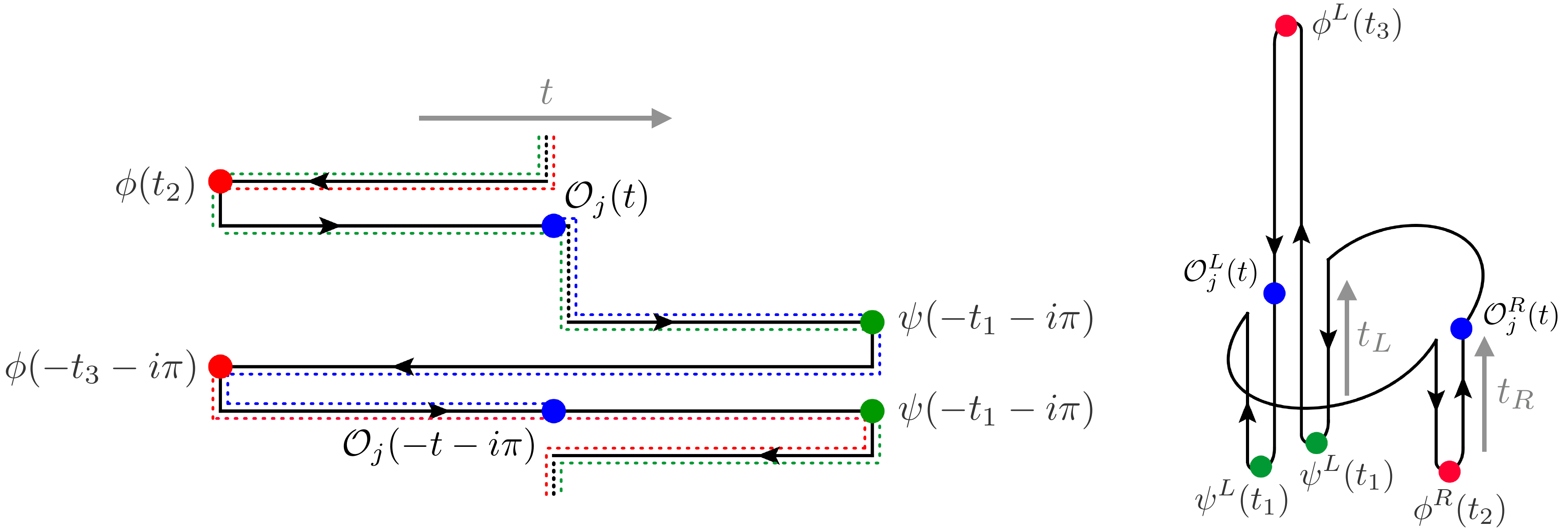}
\end{center}
\vspace{-.4cm}
\caption{Six-point configuration for the traversable wormhole setup. The blue operators correspond to the insertions of ${\cal O}_j^L(t) {\cal O}_j^R(t)$ whose exponential can make the wormhole traversable. The `signal' $\phi$ corresponds to the red insertions. The `perturbation' $\psi^L(t_1)$ is shown in green.} 
\label{fig:contour6foldTraversable}
\end{figure}
In appendix \ref{app:traversable} we compute the `signal' $-2i \text{Im}{\cal C}$. The important feature to note is that this can again be obtained from six-point functions: expanding the operator $e^{-igV}$ in $g$, we can treat the scattering with each factor of $ {\cal O}_j^L {\cal O}_j^R$ as an independent event. Effectively, we are therefore instructed to compute six-point functions of the form 
\begin{equation}
    \big{\langle}  \psi^L(t_1)\phi^L(t_3) \, {\cal O}_j^L(t) {\cal O}_j^R(t)\, \phi^R(t_2) \psi^L(t_1) \big{\rangle}\,.
\end{equation}
This type of six-point function is illustrated in figure \ref{fig:contour6foldTraversable}. It can be computed using similar eikonal resummation techniques as in previous sections (see in particular section \ref{sec:calculations}). After performing some of the eikonal integrals analytically, we find the following expression:
\begin{equation}
\label{eq:traversableRes}
\begin{split}
\cal C &\approx 
   \frac{1}{ \Gamma(2\Delta_\phi)} \, \int_{-\infty}^0 \frac{d\tilde{p}_+}{-p_+} (i\tilde{p}_+)^{2\Delta_\phi} \,\exp \left\{ -i\left(1+ \frac{\Delta_\psi\,e^{-t_1}}{ \sin\delta}\cdot \frac{G}{8}  \frac{e^{\frac{t_3-t_2}{2}}}{\cosh \frac{t_3 + t_2}{2}} \right)\tilde{p}_+ \right\}\\
  &\qquad\qquad\qquad\qquad   \times \exp\left\{-ig \langle V \rangle_\psi \, \qty[1-\qty(1  - \frac{ 1+\frac{G\Delta_{\psi}}{4\sin\delta}\,e^{-t_1} }{2+\frac{G\Delta_{\psi}}{4\sin\delta}\,e^{-t_1}}\cdot\frac{G}{8}  \frac{ e^{\frac{t_3-t_2}{2}}}{\cosh \frac{t_3+t_2}{2}}\,\tilde{p}_+)^{-2\Delta_j}] \right\}
  \end{split}
\end{equation}
where we set $t=0$ for simplicity.
 The momentum integral can in fact be performed analytically in terms of hypergeometric functions, which can be resummed to obtain an analytical approximation to the integral. The resulting expression has a nice physical interpretation as an infinite sum of scattering events involving any number number of particles. Each term in the sum exhibits an elaborate factorization property. For more details, see \eqref{eq:TraversRes}. To extract physical lessons, we will now consider a particular `probe limit' and then analyze the full integral numerically.

\subsection{Probe limit}

A simple way to analyze the integral \eqref{eq:traversableRes} is in the probe limit, where we ignore the backreaction of $\phi$ (i.e., we treat $G \frac{\exp \frac{t_3-t_2}{2}}{ \cosh \frac{t_3+t_2}{2} }\, \tilde{p}_+$ as small). In this case the argument of the second line of \eqref{eq:traversableRes} can be expanded perturbatively. To first nontrivial order this gives:
\begin{equation}
\boxed{\;\;
    {\cal C}^{\text{(probe)}}
    = \left[ \frac{1}{1+\frac{G}{4}\left(\frac{\Delta_\psi\, e^{-t_1}}{2\sin\delta}  -g\Delta_j\, \langle V \rangle_\psi\right)\frac{e^{\frac{t_3-t_2}{2}}}{\cosh \frac{t_3+t_2}{2}}   }\right]^{2\Delta_\phi} 
    \;\;}
\end{equation}
Setting $t_3 = -t_2$ this expression has a pole:
\begin{equation}
\label{eq:pole}
 \text{pole:}\qquad   t_3\equiv-t_2 = -\log \left[ \frac{G}{4} \left(g\Delta_j\,\left[ \frac{1}{2+ \frac{G\Delta_\psi }{4 \sin\delta}\,e^{-t_1} } \right]^{2\Delta_j} - \frac{\Delta_\psi\,e^{-t_1}}{2\sin\delta} \right)\right]\,.
\end{equation}
\begin{figure}
\begin{center}
\includegraphics[width=0.325\textwidth]{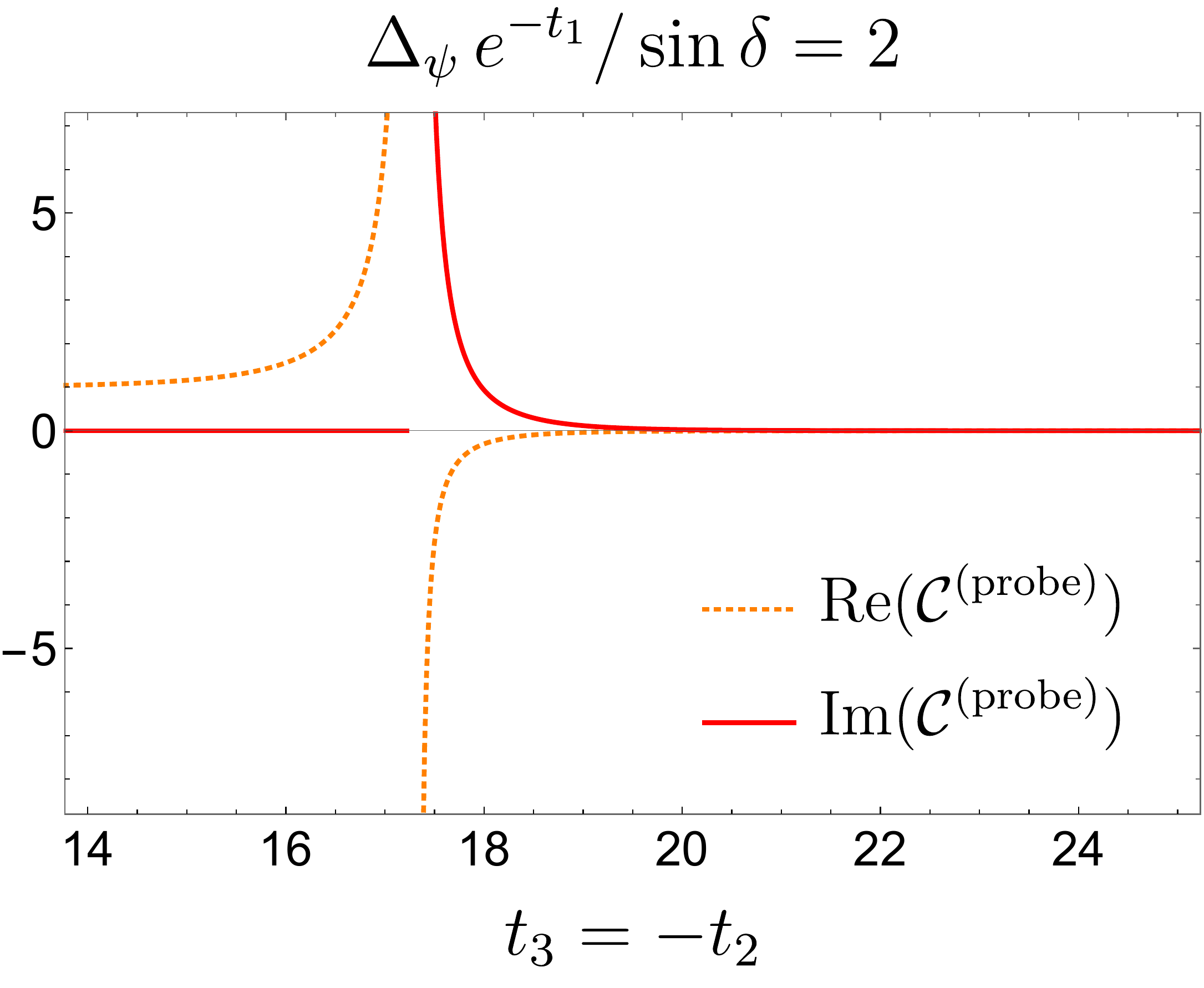}
\includegraphics[width=0.325\textwidth]{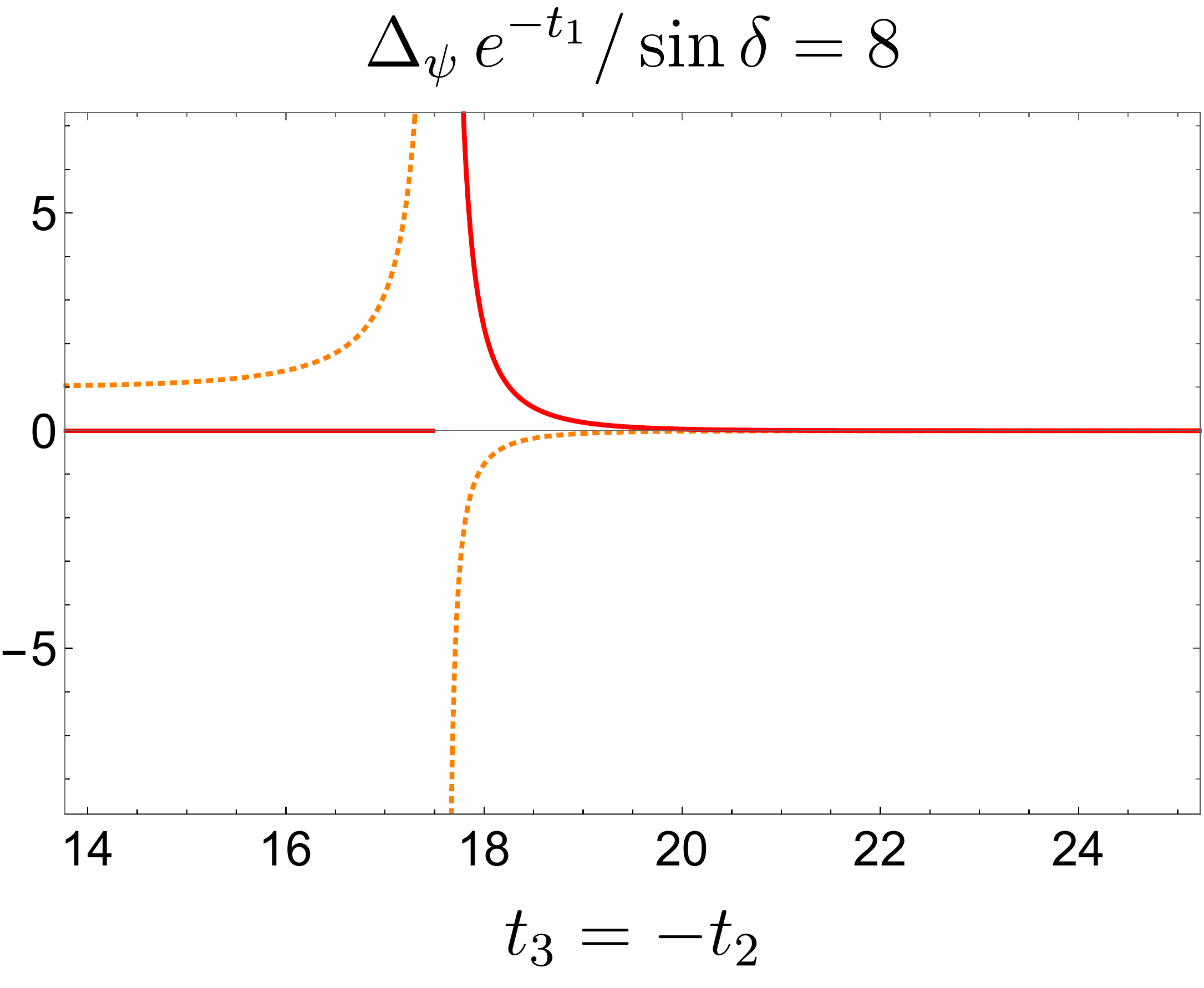}
\includegraphics[width=0.325\textwidth]{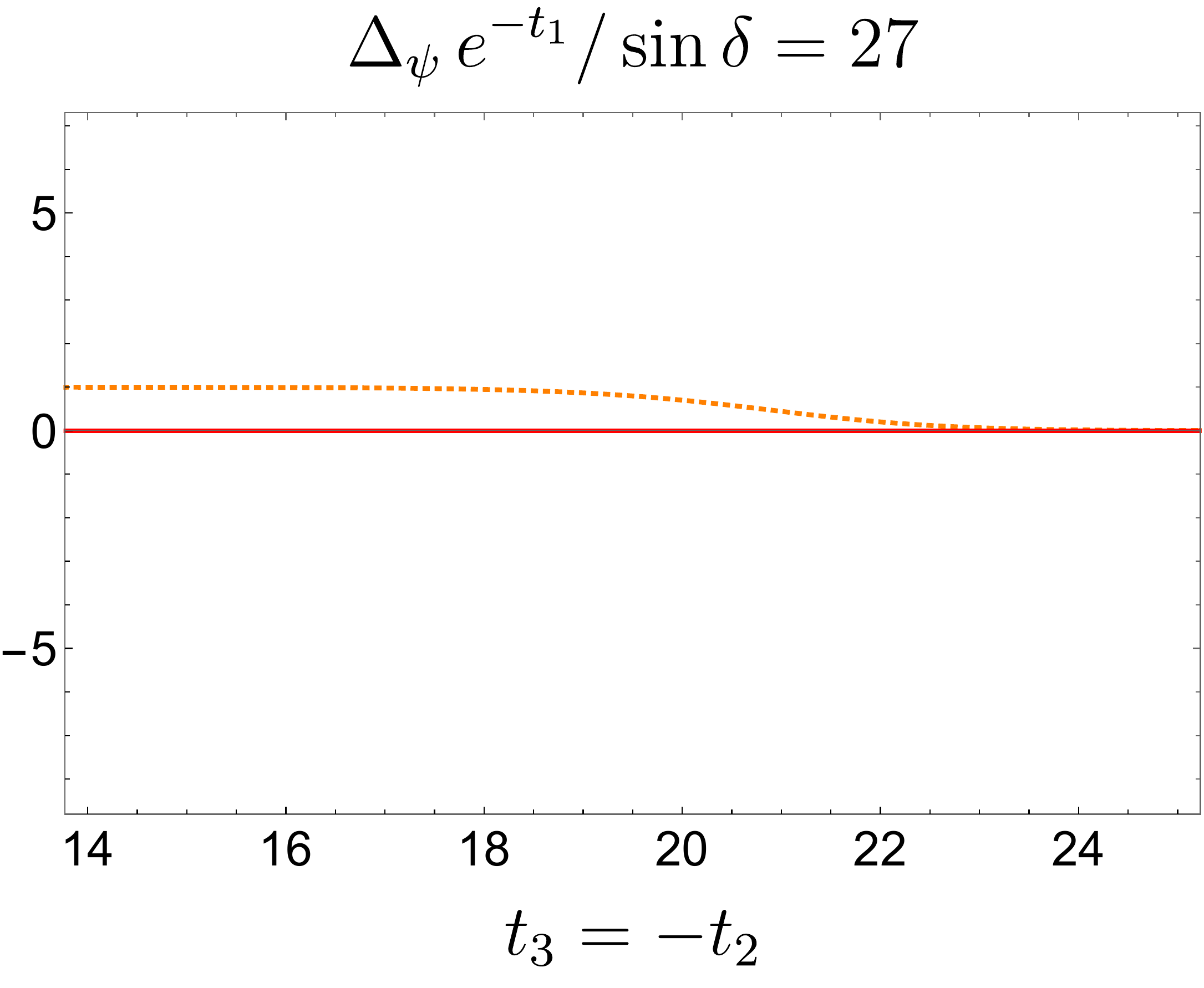}
\end{center}\vspace{-.5cm}
\caption{The signal passing from the right to the left boundary through the traversable wormhole is given by $-2i\, \text{Im}({\cal C})$. This plot shows the probe approximation. We fix the strength of the left-right coupling as $g=50$ and vary the strength of the perturbation, which is parametrized by $\Delta_\psi \, e^{-t_1}/\sin\delta$. For a mild collision (left figure) we observe a pole, indicating that information about the collision has traversed the wormhole. The signal of a strong collision (right figure) cannot be extracted. We set $\Delta_\phi = \Delta_{j} = 0.7$, $G=10^{-8}$. The scrambling time is $t_* \approx -\log G \approx 18$.}
\label{fig:signalsProbe}
\end{figure}
\begin{figure}
\begin{center}
\includegraphics[width=0.325\textwidth]{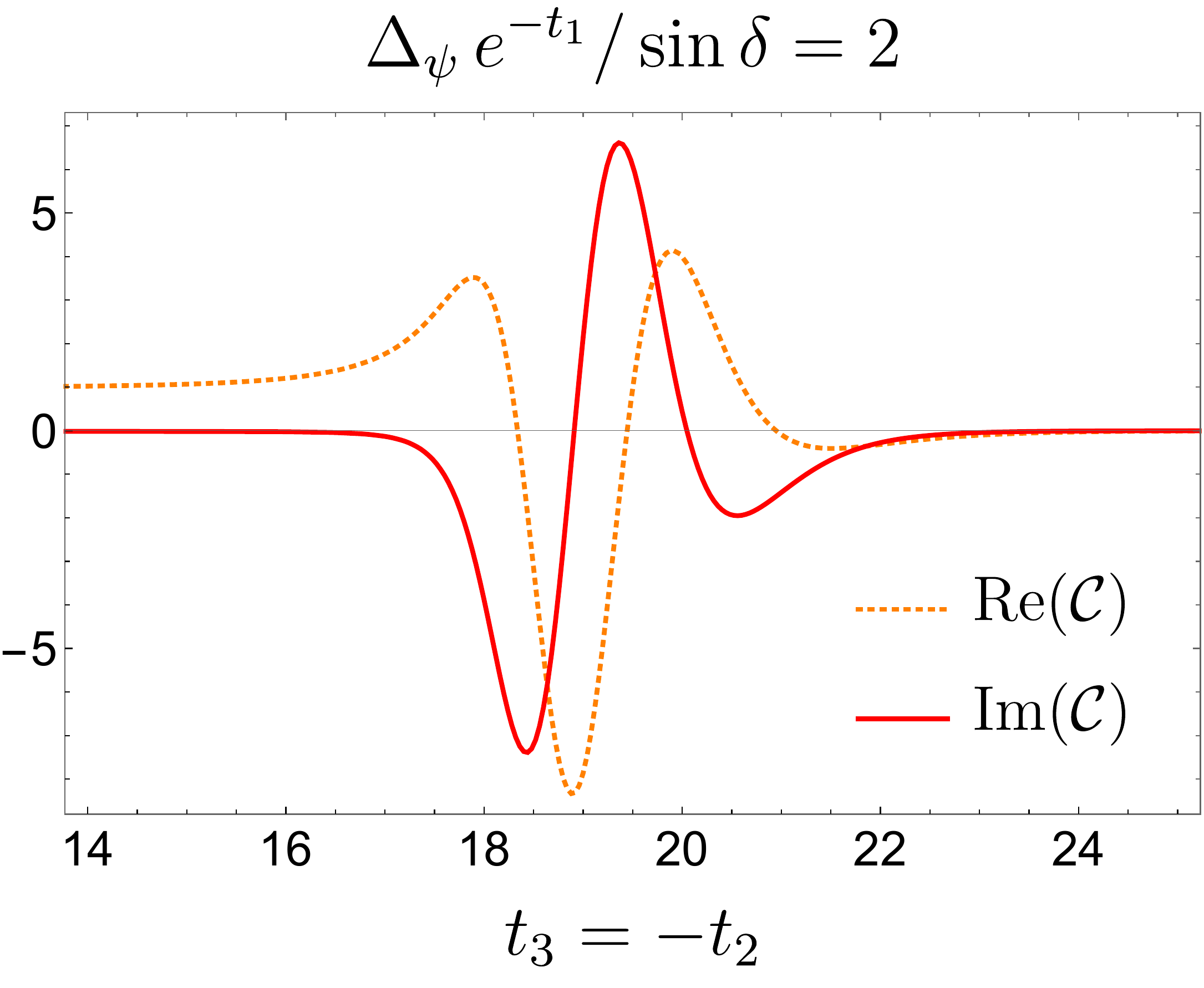}
\includegraphics[width=0.325\textwidth]{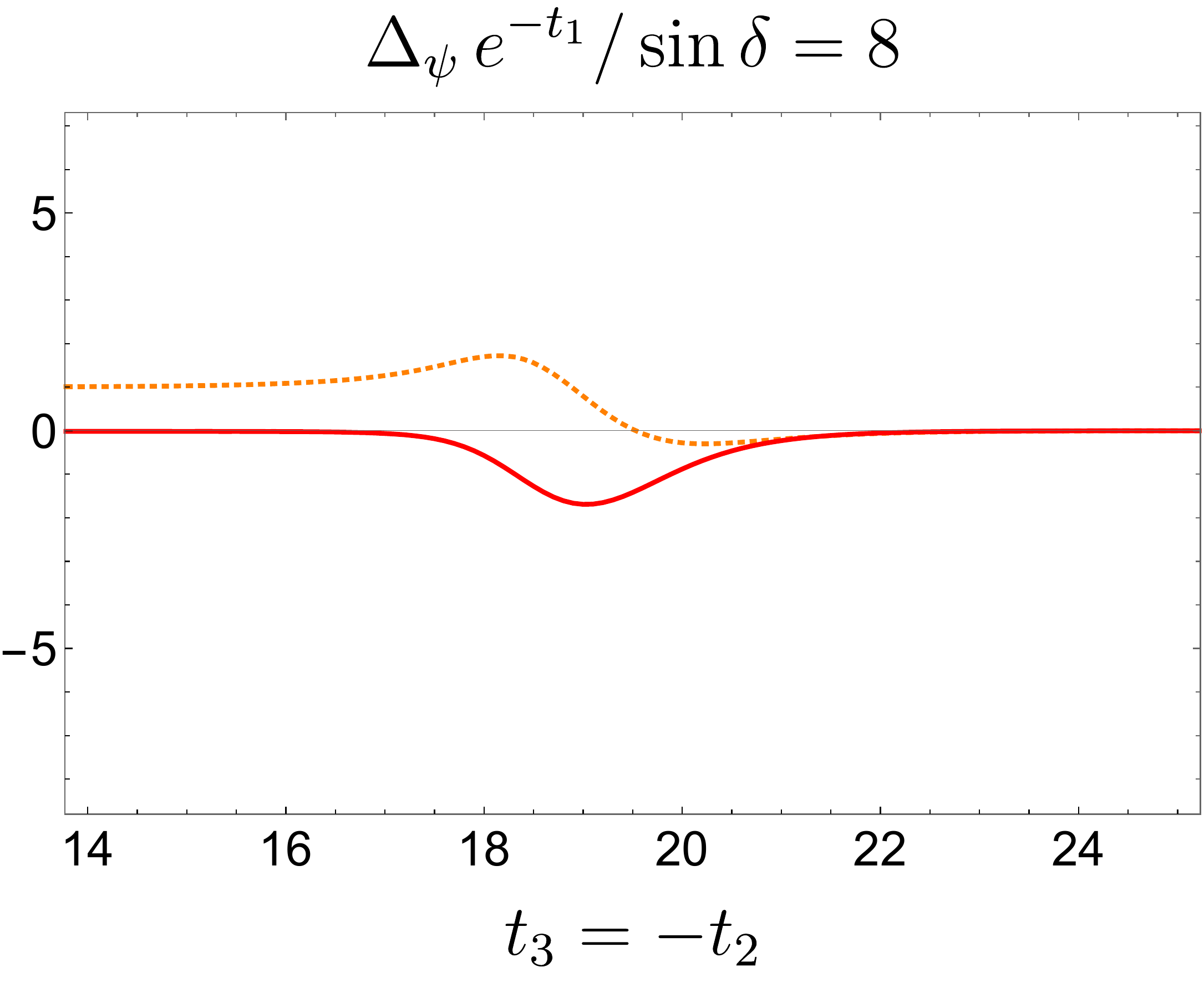}
\includegraphics[width=0.325\textwidth]{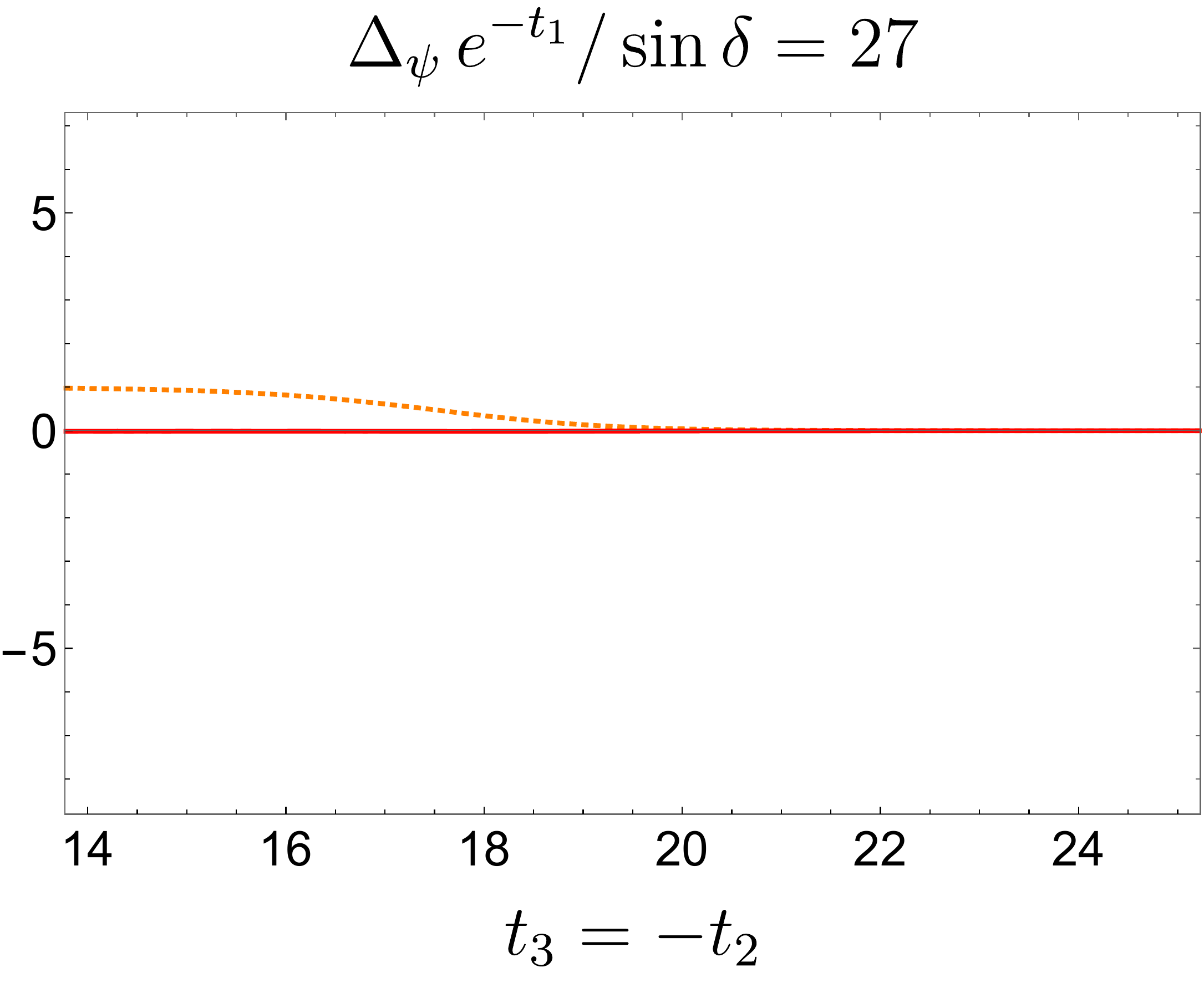}
\end{center}\vspace{-.5cm}
\caption{The numerical results for ${\cal C}$ from \eqref{eq:traversableRes} with the same setup as shown for the probe limit in figure \ref{fig:signalsProbe}. The strength of the perturbation is parametrized by $\Delta_\psi \, e^{-t_1}/\sin\delta$. For a weak perturbation (left figure) the negative energy is enough to make the wormhole traversable and we obtain a clear signal from the collision. In case of a strong collision (right figure) the `teleportation operator' is not strong enough to extract a signal. Parameters: $g=50$, $\Delta_\phi = \Delta_j = 0.7$, $G=10^{-8}$.}
\label{fig:signals}
\end{figure}
This pole is the first indication that a signal has exited the wormhole and reached the left boundary. We can even see a more detailed feature of the exact result already in the probe approximation: the pole disappears if the argument of the logarithm \eqref{eq:pole} becomes negative. This illustrates that the negative energy due to the teleportation operator (measured by $g\Delta_j$) competes with the positive energy of the perturbation $\psi$ (measured by $\Delta_\psi \, e^{-t_1}/\sin\delta$). If the strength of the collision is too large, the latter effect wins, the pole disappears, and the signal has thus not exited the wormhole. We show these features in figure \ref{fig:signalsProbe}. Note that the probe limit cannot distinguish the strength or `quality' of the signal (i.e., the left and middle panels of the figure look identical up to the location of the pole).

\subsection{Numerical result}
\label{sec:numerical}

Figure \ref{fig:signals} shows the real and imaginary parts of the correlator ${\cal C}$, obtained via numerical integration of \eqref{eq:traversableRes}. We fix the strength of the teleportation coupling $g$ such that one expects to be able to teleport a certain amount of information. Indeed, by changing the strength of the collision in the interior (parametrized by the combination $\Delta_\psi\,e^{-t_1}/\sin\delta$), we see qualitatively different behavior: if the collision is weak (left panel of figure \ref{fig:signals}), we obtain a clear signal, indicating that information about the collision product has escaped the black hole. By increasing the collision energy (middle and right panels), the signal disappears: the positive energy shock due to the collision with $\psi$ is now more than the what the teleportation operator with $g=50$ can compensate for. See figure \ref{fig:bulkSetup}(c) for an illustration of this competition. 
Note also that the signal is the strongest when the $\phi^L$ and $\phi^R$ operators are inserted at roughly a scrambling time in the future and past of the teleportation operator $e^{igV}(t=0)$.

In determining whether or not the signal is nontrivial, also the value of $t_1$ matters: if the perturbation $\psi^L(t_1)$ is inserted too far in the past ($-t_1 \sim t_*$), then it will destroy the traversability. This is expected: after a scrambling time, the perturbation has disrupted the entanglement between the left and right boundary theories too much for the teleportation protocol to work.

\section{Eikonal calculations and general observations}
\label{sec:calculations}

In this section we summarize the calculations leading to the results quoted so far. We shall particularly focus on the calculation of ${\cal F}_6$ relevant to sections \ref{sec:collision} and \ref{sec:strength}. Similar calculations regarding the traversable wormhole setup can be found in appendix \ref{app:traversable}.

\begin{figure}
\begin{center}
\includegraphics[width=0.8\textwidth]{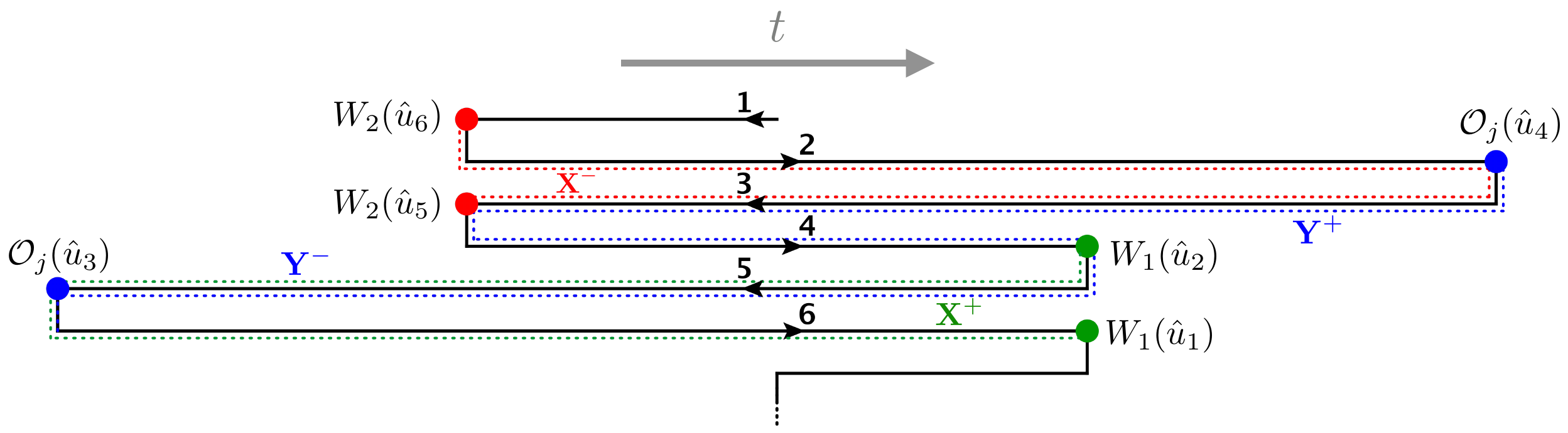}
\end{center}
\vspace{-.6cm}
\caption{Six-point configuration used to diagnose the collision. We indicate the parts of the contour where $SL(2,\mathbb{R})$ parameters $X^\pm$ and $Y^\pm$ have support.} 
\label{fig:contour_largeab}
\end{figure}

To simplify notation, let us write ${\cal F}_6(t_1,t_2;a,b)$ as an analytically continued one-sided correlator:
\begin{equation}
\label{eq:F6def2}
  {\cal F}_6 = \frac{\big{\langle}W_1(\hat{u}_1) {\cal O}_j(\hat{u}_3) W_1 (\hat{u}_2)  W_2(\hat{u}_5){\cal O}_j(\hat{u}_4)  W_2(\hat{u}_6) \big{\rangle}_\beta}{ \langle W_1(\hat{u}_1)W_1(\hat{u}_2)\rangle_\beta\, \langle {\cal O}_j(\hat{u}_3){\cal O}_j(\hat{u}_4)\rangle_\beta \,\langle W_2(\hat{u}_5) W_2(\hat{u}_6)\rangle_\beta}
\end{equation}
with the regularized insertion times
\begin{equation}
\label{eq:ConfigDef2}
 \hat{u}_1 = -t_1- i(\pi+2\delta_1) \,,\;\; \hat{u}_2 = -t_1 -i\pi   \,,\;\; \hat{u}_3= -a-i\pi  \,,\;\; \hat{u}_4 = b\,,\;\; \hat{u}_5 = t_2-2i\delta_2 \,,\;\; \hat{u}_6 = t_2\,.
\end{equation}
The corresponding contour is shown in figure \ref{fig:contour_largeab}.  The right boundary time is related to the complex contour time by $\hat{u} = t_R$. Similarly, the left boundary time is related to the complex contour time by  $\hat{u} = -t_L-i\pi$.

We now discuss how to compute this correlation function using the eikonal approach. Then, we will comment on some of its general properties.

\subsection{Eikonal formalism for the six-point function}
\label{sec:eikonalMain}

As indicated in \eqref{eq:FtildeDef}, we would like to think of the six-point function ${\cal F}_6$ as a two-point correlator in a perturbed thermofield double state. We represent the bra- and ket-states in momentum space by projecting onto tensor product states $|p_+,p_-\rangle$ with definite momentum in null directions:
\begin{equation}
\begin{split}
   W_2^R(t_2) W_1^L(t_1) |\text{TFD}\rangle &= \int dp_+ dp_- \,  |p_+,p_- \rangle \langle p_+ | W_2 \rangle  \langle p_- | W_1 \rangle  \,,\\
   \langle \text{TFD} | W_1^L(t_1) W_2^R(t_2)  &= \int dp_+' dp_-' \,   \langle   W_2 | p_+' \rangle  \langle  W_1| p_-'  \rangle \langle p_+',p_-' | \\
\end{split}
\end{equation}
To compute the two-point function of matter fields sourced by ${\cal O}_j$ operators in the above state, we need to evaluate the projection onto the momentum states.
The eikonal approximation amounts to the statement that these lead to `out-states', which are identical to the original state $|p_+,p_-\rangle$ up to a phase that is induced by the gravitational scattering:
\begin{equation}
\begin{split}
\langle p_+',p_-' | {\cal O}_j^L {\cal O}_j^R |p_+,p_- \rangle
&\approx \delta(p_+' - p_+) \delta(p_-'-p_-)\!  \int dq_+ dq_- \; \langle   {\cal O}_j^L {\cal O}_j^R | q_+,q_- \rangle\,e^{iG (p_+q_- + p_- q_+)} 
\end{split}
\end{equation} 
where $G$ is Newton's constant. Therefore:
\begin{equation}
\label{eq:F6first}
 {\cal F}_6 \approx {\cal N} \int dp_+dp_- dq_+dq_- \, \langle W_2 | p_+ \rangle \langle p_+ | W_2 \rangle\langle W_1 | p_- \rangle \langle p_- | W_1 \rangle \langle   {\cal O}_j^L {\cal O}_j^R | q_+,q_- \rangle\,e^{iG (q_+p_- + q_- p_+)}
\end{equation}
where we remind the reader that ${\cal N}$ denotes the normalization by products of two-point functions.
The eikonal phases account for the high energy scattering of bulk particles with null momenta $p_\pm$ and $q_\mp$. The required wavefunctions such as $\langle p_+ | W_2 \rangle$ etc., are obtained by Fourier transforming bulk-to-boundary propagators along the relevant horizons. Similarly, products of wavefunctions are simply the momentum space representations of boundary correlators subject to finite shifts. For example:
\begin{equation}
\label{eq:waveFc1}
\begin{split}
\langle   W_1 | p_-  \rangle\langle p_- | W_1\rangle
&= \int \frac{dY^-}{2\pi} \, e^{iY^-p_-} \, \big\langle W_1(\hat{u}_2)\, e^{-iY^- \hat{P}_-} \, W_1(\hat{u}_1)\big\rangle \,,\\
\langle   W_2 | p_+  \rangle\langle p_+ | W_2 \rangle
&= \int \frac{dY^+}{2\pi} \, e^{-iY^+p_+} \, \big\langle W_2(\hat{u}_5)\, e^{iY^+ \hat{P}_+} \, W_2(\hat{u}_6) \big\rangle  \,,
\end{split}
\end{equation}
where $\hat{P}_\pm$ are the momentum operators. Similarly,
\begin{align}
\label{eq:waveFc2}
& \langle   {\cal O}_j^L {\cal O}_j^R | q_+,q_- \rangle \nonumber \\
&\quad = 
   \int \frac{dX^+}{2\pi} \frac{dX^-}{2\pi} \, e^{-iX^+ q_+ + iX^- q_-} \,  \Big\langle \!\left( e^{-i X^+ \hat{P}_+} {\cal O}_j(\hat{u}_3)\, e^{i X^+ \hat{P}_+}\right)\!\left( e^{iX^- \hat{P}_-}\,  {\cal O}_j(\hat{u}_4)e^{-iX^- \hat{P}_-} \right)\!
     \Big{\rangle}
\end{align}

In appendix \ref{app:eikonal} we give more details on the explicit form of the momentum space wave functions and their Fourier transforms in the simple case of nearly conformal field theories in one dimension (such as the low-energy sector of the SYK model), see \eqref{eq:BsDef0}, \eqref{eq:BsDef02}.
In any case, combining the above expressions with \eqref{eq:F6first}, the evaluation of ${\cal F}_6$ reduces to an explicit integral over the scalar quantities $X^\pm$ and $Y^\pm$, parametrizing the shifts in the operator insertion points:
\begin{equation}
\label{eq:F6intAbstract}
\begin{split}
{\cal F}_6 &= {\cal N} \int dX^+ dX^- dY^+ dY^- \; \big\langle W_1(\hat{u}_2)\, e^{-iY^- \hat{P}_-} \, W_1(\hat{u}_1) \big\rangle \big\langle W_2(\hat{u}_5)\, e^{iY^+ \hat{P}_+} \, W_2(\hat{u}_6) \big\rangle \\
&\qquad\qquad\qquad\qquad\qquad\qquad \times\Big\langle  
       {\cal O}_j(\hat{u}_3)\, e^{iX^+\hat{P}_+} \; e^{i X^- \hat{P}_- } {\cal O}_j(\hat{u}_4)
     \Big{\rangle} \; e^{iG(X^+Y^- + X^- Y^+ )} 
\end{split}
\end{equation}
This integral can be explicitly evaluated in different ways. We gave the result in previous sections and delegate the detailed calculation to appendix \ref{app:eikonal}.

\subsection{Eikonal method in position space} 
\label{sec:eikonalPosition}

Given the power of the eikonal method as described in the previous subsection, we take the opportunity to offer a different perspective on some of the calculations above. In particular, we would like to understand the eikonal method in position space and in relation to perturbative approaches such as the Schwarzian theory of reparametrization modes. 

In the low-energy SYK model we can perform calculations using the Schwarzian theory \cite{Kitaev:2015aa,Maldacena:2016hyu,Maldacena:2016upp}. The six-point function is then given by an integral over reparametrization fields $\hat{t}(\hat{u})$:\footnote{ Since we will not be computing any quantum corrections, we can ignore the fact that the Schwarzian partition function has a nontrivial measure \cite{Stanford:2017thb}.}
\begin{equation}
\label{eq:F6SchwGen}
  {\cal F}_6 = {\cal N} \int \frac{d\hat{t}[\hat{u}]}{SL(2,\mathbb{R})} \; e^{i S[\hat{t}(\hat{u})]} \, G_{\Delta_1}(\hat{u}_1,\hat{u}_2) \, G_{\Delta_j}(\hat{u}_3,\hat{u}_4) \, G_{\Delta_2} (\hat{u}_5,\hat{u}_6) \,,
\end{equation}
where the normalization ${\cal N}$ is the inverse of the same integral, but without the eikonal phase inserted -- this corresponds to the product of two-point functions as in \eqref{eq:F6def}. The Schwarzian action (in Lorentzian signature) is given by 
\begin{equation}
  iS[\hat{t}(\hat{u})] = -iC \int d\hat{u} \, \{ \hat{t}(\hat{u}),\hat{u} \} \,,
\end{equation}
where the integral runs over real time $\hat{u}$ along contours such as those shown in figures \ref{fig:contour6fold} or \ref{fig:contour6foldSymm}. 
The reparametrization bilocals take the form of reparametrized conformal two-point functions:
\begin{equation}
\label{eq:Bdef}
\begin{split}
 G_{\Delta}(\hat{u}_a,\hat{u}_b) &=   \left[ \frac{-\hat{t}'(\hat{u}_b)\hat{t}'(\hat{u}_b)}{(\hat{t}(\hat{u}_a)-\hat{t}(\hat{u}_b))^2} \right]^{\Delta}
\end{split}
\end{equation}

We now make the following assumption: instead of doing the integral over arbitrary reparametrizations (modulo $SL(2,\mathbb{R})$), we can approximate the effect of the operator insertions by means of finite $SL(2,\mathbb{R})$ transformations with fixed points at asymptotic locations $\hat{u} \rightarrow \pm \infty$. This is implemented through the following piecewise transformation (c.f., \cite{Maldacena:2016upp} for a similar calculation in the case of four-point functions\footnote{ See also \cite{Goel:2018ubv} for a related approach.}):
\begin{equation}
\label{eq:t6Def}
\begin{split}
  \hat{t}(\hat{u}) &= x - \frac{(1+x)^2X^-}{4+(1+x)X^-} \, \theta(2,3) + \frac{(1-x)^2X^+}{4+(1-x)X^+} \, \theta(5,6) \\
 &\qquad + \frac{(1-x)^2Y^+}{4+(1-x)Y^+} \, \theta(3)+ (\cdots) \, \theta(4) - \frac{(1+x)^2Y^-}{4+(1+x)Y^-} \, \theta(5) \,,
 \end{split}
 \end{equation}
where $x = \tanh \frac{\hat{u}}{2}$. The $\theta$-functions indicate the contours on which the respective $SL(2,\mathbb{R})$ transformations have support (c.f., figure \ref{fig:contour_largeab}). The explicit form of the term on the fourth contour will not play any role. 
Because \eqref{eq:t6Def} is {\it not} an $SL(2,\mathbb{R})$ transformation on contours 3 and 5, the Schwarzian action evaluated on this piecewise reparametrization is nontrivial:
\begin{equation}
\label{eq:SchwazianExpand}
   i S[\hat{t}(\hat{u})] =  \frac{iC}{2} \, (X^+ Y^- + X^- Y^+) + \ldots 
\end{equation}
where we work perturbatively in $X^\pm$ and $Y^\pm$ (subsequent terms are parametrically smaller in $\sim \frac{1}{C}$). The bilocal operators \eqref{eq:Bdef} evaluated on the piecewise transformation $\hat{t}(\hat{u})$ become precisely the Fourier transforms of the momentum space wavefunctions, \eqref{eq:BsDef0} and \eqref{eq:BsDef02}:
\begin{equation}
\label{eq:BsDef}
\begin{split}
G_{\Delta_1}(\hat{u}_1,\hat{u}_2) & =  \big\langle W_1(\hat{u}_2)\, e^{-iY^- \hat{P}_-} \, W_1(\hat{u}_1) \big\rangle\\
G_{\Delta_j}(\hat{u}_3,\hat{u}_4) &= \Big\langle  
       {\cal O}_j(\hat{u}_3)\, e^{i X^+ \hat{P}_+ }\; e^{iX^- \hat{P}_- } {\cal O}_j(\hat{u}_4)
     \Big{\rangle}\\
G_{\Delta_2}(\hat{u}_5,\hat{u}_6) &=\big\langle W_2(\hat{u}_5)\, e^{iY^+ \hat{P}_+} \, W_2(\hat{u}_6) \big\rangle 
\end{split}
\end{equation}
Similarly, the Schwarzian integral \eqref{eq:F6SchwGen} becomes an integral over the piecewise $SL(2,\mathbb{R})$ parameters alone, which coincides with the eikonal result \eqref{eq:F6intAbstract}:
\begin{equation}
\label{eq:F6intAbstract2}
  {\cal F}_6 =   \int dX^+ dX^- dY^+ dY^- \; e^{\frac{iC}{2} \, (X^+ Y^- + X^- Y^+)} \; G_{\Delta_1}(\hat{u}_1,\hat{u}_2) \, G_{\Delta_j}(\hat{u}_3,\hat{u}_4) \, G_{\Delta_2} (\hat{u}_5,\hat{u}_6) \,.
\end{equation}

\subsection{The Lorentzian six-point function ${\cal F}_6$ for general insertion times}

Here we take the opportunity to discuss ${\cal F}_6(t_1,t_2;a,b)$ for general $a,b$ (see, e.g., figures \ref{fig:contour6fold} and \ref{fig:contour6foldSymm}). We first discuss the properties of this six-point OTOC and then give its explicit expression resulting from the gravity calculations.

\subsubsection{OTOCology: general properties of the correlator}

Let us study the general properties of ${\cal F}_6(t_1,t_2;a,b)$ as in \eqref{eq:F6def2} for general $a,b$. It has several interesting features as a Lorentzian quantum field theory observable:
\begin{itemize}
\item As long as $t_{1,2} < 0$, $a > t_1$ and $b>t_2$, it is {\it maximally out-of-time-order}, meaning that it cannot be represented on a complex time contour with fewer than three forward-backward segments (which is the maximal number of switchbacks that can occur for six-point functions without introducing redundancies \cite{Haehl:2017qfl,Haehl:2017eob}).
\item For $a<-t_2$ or $b<-t_1$ (as in figure \ref{fig:contour6foldSymm}), ${\cal F}_6$ is not irreducibly out-of-time-order: it receives some `disconnected' four-point contributions, which are out-of-time-order by themselves. In this case, ${\cal F}_6$ is the OTOC that has been previously discussed in the context of quantum chaos \cite{Haehl:2017pak,Haehl:2018izb,Anous:2020vtw}.\footnote{ Note the following interesting feature of the six-point function relevant to the traversable wormhole setup (figure \ref{fig:contour6foldTraversable}): {\it all} of its four-point factors are OTOCs even though the full six-point function is not maximally out-of-time-order.}
\item For $a>-t_2$ and $b>-t_1$ (as in figure \ref{fig:contour6fold}), removing any pair of identical operators from ${\cal F}_6$ leaves behind a four-point function which is in-time-order (i.e., requires only a single switchback in time). The characteristic exponential decay of ${\cal F}_6$ will therefore originate entirely from its `connected' properties as a six-point function and cannot be attributed to `disconnected' four-point contributions. This elucidates the structure \eqref{eq:F6factorize}. We shall refer to this property by saying that ${\cal F}_6$ is {\it irreducibly out-of-time-order}. This distinguishes our observable from six-point functions that have been considered previously in the literature. 
\item Irrespective of the values of real times, the operator ordering along the complex time contour is always {\it maximally braided} \cite{Haehl:2017pak}. This refers to the fact that the pairwise identical operators are all interlinked in Euclidean time, i.e., their supports along contours overlap (as shown by colored lines in figure \ref{fig:contour6fold})
\end{itemize}

\subsubsection{Result from eikonal saddle point and geodesic calculations}

The saddle point approximation to the eikonal calculation (appendix \ref{app:eikonal1}) yields the following result for general $a,b$:
\begin{equation}
\label{generalab}
\begin{split}
{\cal F}_6 
   \approx
\left[ \frac{1}{1+G\Delta_1 \, z_1} \right]^{2\Delta_j} \left[ \frac{1}{1+ G\Delta_2 \, z_2} \right]^{2\Delta_j} 
\left[ \frac{1}{1+ \frac{ G^2\Delta_1\Delta_2 \, z_1 \, z_2 \, e^{a+b}}{ \left( 1+ G\,\Delta_1 \, z_1 \right) \left( 1+ G\,\Delta_2 \, z_2 \right)} }
\right]^{2\Delta_j} 
\qquad (\Delta_{1,2} \gg \Delta_j)
\end{split}
\end{equation}
where
\begin{equation}
  z_1 = \frac{e^{\frac{a-b}{2}-t_1} }{8\, \sin \delta_1 \cosh \frac{a+b}{2}} \,,\qquad
  z_2 = \frac{e^{-\frac{a-b}{2}-t_2} }{8\, \sin \delta_2 \cosh \frac{a+b}{2}}  \,.
\end{equation}
For $a,b \gg -t_{1,2}$ and for $a=b=0$ this reduces to our previous results \eqref{eq:F6saddle1} and \eqref{eq:F6calc2}, respectively.

A similar result is separately obtained through the geodesic approximation in the corresponding double-shock geometry (appendix \ref{app:3D geometry}) for general $a,b$:
\begin{equation}
\begin{split}
{\cal F}_6^\text{(geodesic)}
   \approx
\left[ \frac{\sqrt{1+\frac{\delta S_1}{S}}}{1+\frac{\delta S_1}{S} \, \tilde{z}_1} \right]^{2 \Delta_j} 
\left[ \frac{\sqrt{1+\frac{\delta S_2}{S}}}{1+\frac{\delta S_2}{S} \, \tilde{z}_2} \right]^{2 \Delta_j}
\left[ \frac{1}{1+ \frac{   \left( \frac{\delta S_1}{S} \, \tilde{z}_1 \right) \left( \frac{\delta S_2}{S} \, \tilde{z}_2 \right)\tilde{z}_{12}}{ \left( 1+\frac{\delta S_1}{S} \, \tilde{z}_1 \right) \left( 1+\frac{\delta S_2}{S} \, \tilde{z}_2 \right)} 
}
\right]^{2 \Delta_j} 
\end{split}
\end{equation}
for $\Delta_{1,2} \gg \Delta_j \gg 1$, where
\begin{equation}
  \tilde{z}_1 = \frac{\cosh\frac{a-t_1}{2} \cosh\frac{b+t_1}{2}}{\cosh\frac{a+b}{2}} \,,\quad
  \tilde{z}_2 = \frac{\cosh\frac{b-t_2}{2} \cosh\frac{a+t_2}{2}}{\cosh\frac{a+b}{2}} \,,\quad
  \tilde{z}_{12} = \frac{\sinh\frac{a-t_1}{2} \sinh\frac{b-t_2}{2}}{\cosh\frac{b+t_1}{2}\cosh\frac{a+t_2}{2}}
  \,.
\end{equation}
and $\delta S_i$ is the entropy introduced by the respective operator insertion, which scales as $G \Delta_i$ (see \ref{eq:SDelta}). Note that if we ignore the terms proportional to $\frac{\delta S_i}{S}$ but not enhanced by $e^{-u_i}$, this reduces to \eqref{generalab}. See also \eqref{eq:F6calc2appendix} and \eqref{eq:JTfinalResult}.  In particular, for $a,b \gg -t_{1,2}$ and for $a=b=0$ this reduces to our previous results \eqref{collision_diagnosis} and \eqref{eq:F6calc2}, respectively.  

Regarding the gravitational calculation, a few comments are in order. The factorization property of the $\mathcal{F}_6$ correlator has an alternative interpretation from this geometric perspective. In the geodesic approximation, the correlator decays with the exponential of the geodesic distance between the two $\mathcal{O}_j$ probes. Since the state has been perturbed by the $W_1, W_2$ insertions, the corresponding geometry is now that of the background with two additional shocks. Consequently, the resultant distance naturally expands into a sum of the background, the independent effects of the $W_1, W_2$ insertions, and one ``connected" contribution. Such an expansion of distance effects then exponentiates into the factorization of the resultant correlator. This connected contribution is due to the fact that gravity is not linear, and it corrects for the fact that the two insertions affect each other.

\section{Conclusion}
\label{sec:conclusion}
Using out-of-time-order six-point functions, we quantified various properties of collisions behind the horizon of particles sent into the eternal AdS black hole from the two asymptotic regions. In sections \ref{sec:collision}, \ref{sec:strength} and \ref{sec:traversable}, we considered three different real-time configurations to elucidate different aspects of the collision. At a technical level, we used the eikonal resummation technique as well as gravitational geodesic computations to obtain the relevant expressions for the six-point functions. A common theme was the factorization property of the eikonal six-point function into four-point factors and a connected contribution, c.f., \eqref{eq:F6factorize}.\\

\noindent Some unanswered questions are as follows:
\begin{itemize}
\item In the setup for diagnosing the collision in the wormhole interior, it appears that the existence of the singularity plays a crucial role by means of prohibiting the meeting of signals that are sent in too late. What will happen for geometries without a singularity, such as charged black holes? In that case, signals can meet no matter how late they are sent in, but they will meet in a region where the dilaton is negative, or in other words, where gravity is repulsive. It will be important to understand this phenomenon in the future as the six-point functions we used are not sensitive to this. 
\item In the setup for detecting the overlap of two perturbations spreading through a shared quantum circuit, 
the connected part of the six-point function \eqref{eq:F6calc2} gives an order one contribution. This is qualitatively negligible, but physically interesting as it quantifies the amount by which the perturbations do {\it not} spread independently. Is there any way to account for this factor in the circuit model?
\item In the traversable wormhole setup, we can only extract the signal after the collision if the collision is mild. It seems that this protocol does not allow us to extract information about a strong collision. Similarly, we only demonstrated how to extract one collision product (the $\phi$ signal), but not the other one (the $\psi$ signal). These are deficiencies of our protocol. Are there other protocols to improve the situation?\footnote{For recent work in this direction, see \cite{Lensky:2020fqf}.} 
\item The factorization property \eqref{eq:F6factorize} of the eikonal six-point function calls for a more detailed understanding. Similar structures have appeared before in two-dimensional CFTs \cite{Fitzpatrick:2015qma,Anous:2019yku,Anous:2020vtw}, where the eikonalization of Virasoro conformal blocks offers an explanation. It would be interesting to argue for \eqref{eq:F6factorize} in greater generality (perhaps using a perturbative expansion as in appendix \ref{sec:perturbative}). This would open up the possibility to understand higher-point Virasoro identity blocks in interesting regimes such as the Regge limit or in configurations with several heavy operators (generalizing \cite{Fitzpatrick:2015dlt}), which would have many applications. We expect that many of the expressions derived in this paper have a direct analog in Virasoro identity conformal blocks.
\end{itemize}

\section*{Acknowledgements}
We thank Ahmed Almheiri, Tarek Anous, Adam Levine, Henry Lin, Juan Maldacena for helpful discussions. A.S. was supported in part by the Perimeter Institute. A.S.\ and Y.Z.\ are supported by the Simons foundation through the It from Qubit Collaboration. F.H.\ is grateful for support from the DOE grant DE-SC0009988 and from the Paul Dirac and Sivian Funds.

\appendix 

\section{Eikonal calculations in the Schwarzian theory}
\label{app:eikonal}

In this appendix we give some details regarding the evaluation of the eikonal integrals \eqref{eq:F6intAbstract} and a similar expression for the traversable wormhole setup. 

\subsection{Eikonal calculation of ${\cal F}_6$}
\label{app:eikonal1}

\paragraph{Momentum generators:}
We define AdS$_2$ in terms of embedding space coordinates $Y \equiv (Y^{-1},Y^0,Y^1) \in \mathbb{R}^{2|1}$ as the universal cover of the surface $Y^2 = -1$. We also work with lightcone coordinates $Y^\pm \equiv Y^0 \pm Y^1$. The embedding space metric is simply $\eta = \text{diag}(-1,-1,1)$. Points $X$ on the boundary are defined up to the equivalence relation $X \sim \lambda X$ and satisfy $X^2=0$.
The shockwave scattering in the bulk induces shifts described by the translation generators $\hat{P}_\pm$. These act as follows on the boundary points $X_{R,L}$ at the right and left boundaries:
\begin{equation}
\begin{split}
     &e^{-ia^+P_+}: \qquad \qty(Y^{-1}, Y^+, Y^-)\longrightarrow 
	\qty(Y^{-1}-\frac{a^+}{2}Y^-, \;Y^++a^+Y^{-1}-\frac{(a^+)^2}{4}Y^-,\; Y^-)\\
     &e^{-ia^-P_-}:\qquad \qty(Y^{-1}, Y^+, Y^-) \longrightarrow 
	\qty(Y^{-1}-\frac{a^-}{2}Y^+, \;Y^+, \;Y^-+a^-Y^{-1}-\frac{(a^-)^2}{4}Y^+)
\end{split}
\end{equation}
Left and right operator insertions are parameterized as $X_L(t_L) = (1,-e^{-t_L},e^{t_L})$ and $X_R(t_R) = (1, e^{t_R} , -e^{-t_R})$, respectively. 
For instance, the boundary insertion points of the six operators in \eqref{eq:F6def2} are
\begin{equation}
\begin{split}
W_1^L(t_1): \quad X_L^{(W_1)}(t_1) &= (1,\, -e^{-t_1},\, e^{t_1} )\,,\\
W_2^R(t_2): \quad X_R^{(W_2)}(t_2) &= (1,\, e^{t_2},\, -e^{-t_2} )\,,\\
 {\cal O}_j^L(A), \,{\cal O}_j^R(b): \quad X_L^{({\cal O}_j)} (a) &= (1,\, -e^{-a},\,e^a ) \,,\quad X_R^{({\cal O}_j)}(b) = (1,\, e^b,\,-e^{-b} )  \,.
\end{split}
\end{equation}
Acting with the momentum operators on conformal two-point functions such as
\begin{equation}
  \langle W_1(\hat{u}_1) W_1(\hat{u}_2) \rangle = \left[ -2 \, X^{(W_1)}(\hat{u}_1) \cdot X^{(W_1)}(\hat{u}_2) \right]^{-\Delta_1} = \left[ \frac{-i}{2\sinh \frac{\hat{u}_{12}}{2}} \right]^{2\Delta_1} \,,
\end{equation}
we immediately obtain the expressions \eqref{eq:BsDef0} and \eqref{eq:BsDef02}, which are also reparametrized two-point functions \eqref{eq:BsDef}.

\paragraph{Explicit wave functions in NCFT${}_1$:}
The position space wave functions \eqref{eq:waveFc1} and \eqref{eq:waveFc2} are rather general, but can be evaluated very explicitly in nearly conformal field theories in one dimension (or in higher-dimensional CFTs): they are just two-point functions, which are determined by symmetry in the usual way, and the momentum generators act by simple translations. We find, for example:
\begin{equation}
\label{eq:BsDef0}
\begin{split}
\big\langle W_1(\hat{u}_2)\, e^{-iY^- \hat{P}_-} \, W_1(\hat{u}_1)\big\rangle  & =   \left[ \frac{i}{ 2\sinh \frac{\hat{u}_{21}}{2} - \frac{Y^{^-}}{2} e^{\frac{\hat{u}_1+\hat{u}_2}{2}}}  \right]^{2 \Delta_1} 
 =  \left[ \frac{i}{ 2i\sin \delta_1 +  \frac{Y^{^-}}{2}  e^{-t_1 - i\delta_1}}  \right]^{2 \Delta_1}\\
\big\langle W_2(\hat{u}_5)\, e^{iY^+ \hat{P}_+} \, W_2(\hat{u}_6) \big\rangle  &
 =   \left[ \frac{-i}{ 2\sinh \frac{\hat{u}_{56}}{2} + \frac{Y^{^+}}{2}  e^{-\frac{\hat{u}_5+\hat{u}_6}{2}}}  \right]^{2 \Delta_2} 
 =\left[ \frac{i}{2i\sin \delta_2 - \frac{Y^{^+}}{2} e^{-t_2+i\delta_2}}  \right]^{2 \Delta_2}
\end{split}
\end{equation}
and 
\begin{equation}
\label{eq:BsDef02}
\begin{split}
&\Big\langle  
       {\cal O}_j(\hat{u}_3)\, e^{iX^+\hat{P}_+} \; e^{i X^- \hat{P}_- } {\cal O}_j(\hat{u}_4)
     \Big{\rangle}  = \left[ \frac{1}{2\cosh \frac{a+b}{2}  +\frac{X^{^-}}{2} \, e^{-\frac{a-b}{2} 
 }- \frac{X^{^+}}{2} \, e^{\frac{a-b}{2} } - \frac{X^{^+}X^{^-}}{4}\, e^{\frac{a+b}{2}}} \right]^{2\Delta_j} \end{split}
\end{equation}

\paragraph{Fourier transforms:}
Next, we compute the Fourier transforms of the bilocal operators \eqref{eq:BsDef0} and \eqref{eq:BsDef02}, following \cite{Shenker:2014cwa,Maldacena:2017axo}. For instance, we define
\begin{equation}
\begin{split}
G_{\Delta_1} (\hat{u}_1,\hat{u}_2) \equiv \big\langle W_1(\hat{u}_1)\, e^{-iY^- \hat{P}_-} \, W_1(\hat{u}_2) \big\rangle &\equiv  \int dp_- \; e^{-iY^- p_-} \, \langle W_1 | p_- \rangle \langle p_- | W_1 \rangle \,.
\end{split}
\end{equation}
Using the explicit expression \eqref{eq:BsDef0}, we find the wavefunction by Fourier transform:
\begin{equation}
\begin{split}
&\langle W_1 | p_- \rangle \langle p_- | W_1 \rangle
    = \frac{1}{\Gamma(2\Delta_1)} \, \Theta(-p_-) \frac{(-2p_- \, e^{t_1})^{2\Delta_1}}{-p_-} \, \exp \left[ 4p_- \,e^{t_1}\sin\delta_1\right] 
\end{split}
\end{equation}
Note that the exponent has a small imaginary part due to the regulator $\delta_1$, which has the correct sign to make the integral convergent.
 Similarly, we find for the Fourier transform of $G_{\Delta_2}$:
\begin{equation}
\begin{split}
&\langle W_2 | p_+ \rangle \langle p_+ | W_2 \rangle
 =  \frac{1}{\Gamma(2\Delta_2)} \, \Theta(-p_+) \frac{( -2p_+ \, e^{t_2})^{2\Delta_2}}{-p_+}  \, \exp \left[  4p_+ \,e^{t_2}\sin\delta_2\right] 
\end{split}
\end{equation}
Using the two equations above, we can now write the eikonal six-point correlator \eqref{eq:F6first} as follows ($\tilde{p}_\pm = 4p_\pm e^{t_{2,1}} \sin \delta_{2,1}$):
{\small
\begin{equation}
\label{eq:F6calc1}
\begin{split}
{\cal F}_6 
  &= {\cal N} \int dp_+ dp_- \; \langle W_2 | p_+ \rangle \langle p_+ | W_2 \rangle\langle W_1 | p_- \rangle \langle p_- | W_1 \rangle \; \Big\langle  
       {\cal O}_j(\hat{u}_3)\, e^{iG p_- \hat{P}_+ } \; e^{-iG p_+  \hat{P}_- } {\cal O}_j(\hat{u}_4)
     \Big{\rangle} \\
   &=  \frac{1 }{\Gamma(2\Delta_1)\Gamma(2\Delta_2)}  \int_0^{\infty} \frac{d\tilde{p}_+ }{\tilde{p}_+} \; \tilde{p}_+^{2\Delta_2}\int_0^\infty \frac{d\tilde{p}_-}{p_-} \; \tilde{p}_-^{2\Delta_1} \; e^{- \tilde{p}_- \, -\tilde{p}_+} \,   \\
 &\qquad\qquad \times  
 \left[1 + \frac{ \frac{G\tilde{p}_-}{16\sin\delta_1} \, e^{\frac{a-b}{2}-t_1} + \frac{ G\tilde{p}_+}{16\sin\delta_2}\, e^{-\frac{a-b}{2}-t_2} +\frac{G^2\tilde{p}_-  \tilde{p}_+}{128 \sin\delta_1 \sin \delta_2}\, e^{\frac{a+b}{2}-t_1-t_2}}{\cosh \frac{a+b}{2}}  \right]^{-2\Delta_j} 
\end{split}
\end{equation}
}\normalsize
where we first did the integrals over $Y^\pm$ (yielding two delta-functions), and then did the trivial $X^\pm$ integrals. Also note that the normalization factor  
\begin{equation}
 {\cal N} = \left[ \frac{1}{2\cosh \frac{a+b}{2}} \right]^{2\Delta_j} \left[ \frac{1}{2\sin \delta_1 }\right]^{2\Delta_1}\left[ \frac{1}{2\sin \delta_2 }\right]^{2\Delta_2} 
\end{equation}
has cancelled completely in the second step above.

\paragraph{Saddle point approximation:}
The most straightforward way to perform the integrals \eqref{eq:F6calc1} is by using a saddle point approximation: when $\Delta_{1,2} \gg \Delta_j$, the integrals are dominated around $\tilde{p}_\pm \approx  2\Delta_{2,1}$ and we immediately get:\footnote{ The overall prefactor being 1 follows from a standard Gaussian saddle point integral and use of the Stirling formula $\Gamma(2\Delta_1) \approx \sqrt{2\pi(2\Delta_1 - 1)} \, (2\Delta_1-1)^{2\Delta_1-1}  \, e^{1-2\Delta_1}$ for large $\Delta_1$ (similarly for $\Delta_2$).}
\begin{equation}
\label{eq:F6calc2appendix}
\begin{split}
{\cal F}_6 
   \approx
\left[ \frac{1}{1+G\Delta_1 \, z_1} \right]^{2\Delta_j} \left[ \frac{1}{1+ G\Delta_2 \, z_2} \right]^{2\Delta_j} 
\left[ \frac{1}{1+ \frac{ G^2\Delta_1\Delta_2 \, z_1 \, z_2 \, e^{a+b}}{ \left( 1+ G\,\Delta_1 \, z_1 \right) \left( 1+ G\,\Delta_2 \, z_2 \right)} }
\right]^{2\Delta_j} 
\qquad (\Delta_{1,2} \gg \Delta_j)
\end{split}
\end{equation}
where
\begin{equation}
\label{eq:z12pdef}
  z_1 = \frac{e^{\frac{a-b}{2}-t_1} }{8\, \sin \delta_1 \cosh \frac{a+b}{2}} \,,\qquad
  z_2 = \frac{e^{-\frac{a-b}{2}-t_2} }{8\, \sin \delta_2 \cosh \frac{a+b}{2}}  \,.
\end{equation}
For $a,b\gg -t_{1,2}$ only the third (`connected') factor matters and we obtain \eqref{eq:F6saddle1}. Similarly, for $a=b=0$ we get \eqref{eq:F6calc2}.

\paragraph{Exact eikonal integral for $a,b \gg -t_{1,2}$:}
Let us now assume $a \approx b \gg -t_{1,2}$ and perform the integral \eqref{eq:F6calc1} without resorting to saddle point approximation. We have:
\begin{equation}
\begin{split}
  {\cal F}_6 &\approx \frac{1}{\Gamma(2\Delta_1)\Gamma(2\Delta_2)}    \int_0^\infty \frac{d\tilde{p}_-}{\tilde{p}_-} \; \tilde{p}_-^{2\Delta_1}\, e^{ -\tilde{p}_- \, }\int_0^{\infty} \frac{d\tilde{p}_+}{\tilde{p}_+} \; \frac{\tilde{p}_+^{2\Delta_2}e^{-\tilde{p}_+ \, }}{ \left[1 +\frac{G^2\,\tilde{p}_-  \tilde{p}_+}{64\sin\delta_1\sin\delta_2}\, e^{-(t_1+t_2)}  \right]^{2\Delta_j}} 
\end{split}
\end{equation}
The integral over $\tilde{p}_+$ can be performed explicitly in terms of the confluent hypergeometric function,
\begin{equation}
  U(\alpha_1,\alpha_2,z) = \frac{1}{\Gamma(\alpha_1)} \int_0^\infty dt \, e^{-zt} \,t^{\alpha_1-1}(1+t)^{\alpha_2-\alpha_1-1} \qquad (\text{Re}(\alpha_1) > 0).
\end{equation} 
We find:
\begin{equation}
\label{eq:F6calc3}
\begin{split}
{\cal F}_6 
 &\approx   \frac{1}{\Gamma(2\Delta_1)} \,\left(\frac{64\,\sin\delta_1\sin\delta_2}{G^2} \; e^{t_1+t_2} \right)^{2\Delta_2} 
 \\
 &\quad\times \int_0^\infty \frac{d\tilde{p}_- }{p_-} \,\tilde{p}_-^{2\Delta_{12}} \; e^{-\tilde{p}_- \, } \,
U\left(2\Delta_2,\, 1+2\Delta_{2j}, \, -\frac{64 \sin\delta_1\sin\delta_2}{G^2} \, \frac{e^{(t_1+t_2) }}{\tilde{p}_-}  \right)
\end{split}
\end{equation}
where $\Delta_{ab} \equiv \Delta_a - \Delta_b$.
This leaves us with the integral over $\tilde{p}_-$, which can also be evaluated explicitly.  If $\Delta_1 = \Delta_2 \equiv \Delta$, we get a compact expression in terms of the Meijer G-function:
\begin{equation}
\label{eq:F6calc4}
\begin{split}
{\cal F}_6 
&\approx   \frac{1}{\Gamma(2\Delta)^2 \Gamma(2\Delta_j)}  \, 
\frac{1}{z^{4\Delta_j}} \;
 G^{31}_{13} \left(\frac{1}{z^2}  \, \Big| \begin{smallmatrix}   1-2\Delta_j   \\ 0 ,\, 2(\Delta-\Delta_j),\, 2(\Delta-\Delta_j) \end{smallmatrix} \right)
\end{split}
\end{equation}
More generally, when all operator dimensions are different, we find \eqref{eq:F6calc5}. These expressions and their formal properties bear striking similarity with four-point results in the Schwarzian theory \cite{Maldacena:2016upp,Mertens:2017mtv,Lam:2018pvp} and in two-dimensional CFTs \cite{Chen:2016cms}. On the other hand, the characteristic time scale for the exponential decay is given by twice the scrambling time, thus distinguishing these observables from the four-point case.

\subsection{Eikonal calculation of ${\cal C}$}
\label{app:traversable}

In this section compute the `signal' $2i \,\text{Im}({\cal C})$ discussed in section \ref{sec:traversable}, which
indicates that the wormhole has become traversable and information about the collision has managed to escape. The basic quantity of interest is ${\cal C}$, which we will compute using the eikonal method. It will be convenient to work with the unnormalized expression $\tilde{\cal C}$ defined through
\begin{align}
\label{eq:CCtilde}
    \mathcal{C} = \frac{e^{-ig\expval{V}_{\psi}}}{\langle \phi^R\phi^L \rangle\langle \psi^L\psi^L \rangle}\;\mathcal{\tilde{C}}\qquad \text{with}\qquad
   \mathcal{ \tilde C} = \big{\langle}\psi^L(t_1)\phi^L(t_3)e^{igV}\phi^R(t_2)\psi^L(t_1)\big{\rangle}\,.
\end{align}
Furthermore, we consider the scattering events involving $n$ copies of $V = {\cal O}^L {\cal O}^R$ as independent, so that 
\begin{equation}
\label{eq:calCdef}
\begin{split}
\tilde{\cal C} &= \sum_{n\geq 0} \frac{(ig)^n}{n!} \, 
\int dp_+dp_-  \, \langle \phi^L | p_+ \rangle \langle p_+ | \phi^R \rangle\langle \psi^L | p_- \rangle \langle p_- | \psi^L \rangle\\
&\qquad\qquad\qquad\qquad \times \left[ \int dq_+dq_- \,    \langle q_+,q_-| \mathcal{O}_j^L{\cal O}_j^R \rangle\,e^{iG (q_+p_- + q_- p_++p_+ p_-)} \right]^n
\end{split}
\end{equation}
where the wave functions are:
\begin{equation}
\label{eq:waveFcsTrav}
\begin{split}
&\langle \phi^L(t_3) | p_+ \rangle \langle p_+ | \phi^R(t_2) \rangle
= \int \frac{dY^+}{2\pi} \, e^{-iY^+p_+} \, \big\langle \phi^L(t_3)\, e^{iY^+ \hat{P}_+} \, \phi^R(t_2) \big\rangle   \,,\\
&\langle \psi^L(t_1) | p_- \rangle \langle p_- | \psi^L(t_1) \rangle  
= \int \frac{dY^-}{2\pi} \, e^{iY^-p_-} \,  \big\langle \psi^L(t_1)\, e^{-iY^- \hat{P}_-} \, \psi^L(t_1-2i\delta) \big\rangle \,,\\ 
& \langle q_+,q_-| \mathcal{O}_j^L{\cal O}_j^R \rangle = 
   \int \frac{dX^+}{2\pi} \frac{dX^-}{2\pi} \, e^{-iX^- q_- + iX^+ q_+} \, \Big\langle  
      e^{iX^- \hat{P}_- } {\cal O}_j^L(t)\, e^{-iX^- \hat{P}_- } \; e^{-i X^+ \hat{P}_+ } {\cal O}_j^R(t)e^{i X^+ \hat{P}_+ }
     \Big{\rangle} 
\end{split}
\end{equation}
In the eikonal integral \eqref{eq:calCdef}, let us first consider a single term in the expansion of the exponential:
\begin{equation}
\label{eq:calC1}
\begin{split}
\tilde{\cal C}_1 \equiv \int dp_+dp_- dq_+dq_- \, \langle \phi^L | p_+ \rangle \langle p_+ | \phi^R \rangle\langle \psi^L | p_- \rangle \langle p_- | \psi^L \rangle   \langle q_+,q_-| \mathcal{O}_j^L{\cal O}_j^R \rangle\,e^{iG (q_+p_- +p_+ p_-+p_+ q_-))}
\end{split}
\end{equation}
The wave functions appearing in \eqref{eq:waveFcsTrav} are explicitly given by:
\begin{equation}
\label{eq:waveFcs8}
\begin{split}
 \big\langle \phi^L(t_3)\, e^{iY^+ \hat{P}_+} \, \phi^R(t_2) \big\rangle 
 &= \left[ \frac{1}{ 2\cosh \frac{t_2+t_3}{2} -\frac{Y^+}{2} \, e^{\frac{t_3-t_2}{2}}+ i\varepsilon}  \right]^{2\Delta_\phi}\,,\\
  \big\langle \psi^L(t_1)\, e^{-iY^- \hat{P}_-} \, \psi^L(t_1-2i\delta) \big\rangle  
  &= \left[ \frac{i}{2i\sin \delta + \frac{Y^-}{2} \, e^{-t_1 }} \right]^{2\Delta_\psi}\,,\\
 \Big\langle  
       {\cal O}_j^L(t)\, e^{-iX^- \hat{P}_- } \; e^{-i X^+ \hat{P}_+ } {\cal O}_j^R(t)
     \Big{\rangle} &= \left[ \frac{1}{2\cosh t  -\frac{X^{^-}}{2} +\frac{X^{^+}}{2}  - \frac{X^{^+} X^{^-}}{4} \, e^{-t}} \right]^{2\Delta_j} 
 \end{split}
\end{equation}
Plugging in these explicit expressions, we find for \eqref{eq:calC1}:
{\small
\begin{equation}
\label{eq:C1intRes}
\begin{split}
\tilde{\cal C}_1 = \int dp_+dp_- \,e^{iG p_+ p_-}\, \langle \phi^L | p_+ \rangle \langle p_+ | \phi^R \rangle\langle \psi^L | p_- \rangle \langle p_- | \psi^L \rangle  \, \left[ \frac{1}{2\cosh t  - \frac{G}{2} (p_++p_-)  + \frac{ G^2}{4}\, p_+ p_-  \, e^{-t}} \right]^{2\Delta_j} 
\end{split}
\end{equation}
}\normalsize
One can also do this computation using the position space method of section \ref{sec:eikonalPosition}, and obtain exactly the same result (c.f. figure \ref{fig:contour6foldTraversable}).

In order to finally evaluate the eikonal integral explicitly, first note the Fourier transforms of the wave functions \eqref{eq:waveFcs8}:
\begin{equation}
\begin{split}
\langle \phi^L | p_+ \rangle \langle p_+ | \phi^R \rangle
&=  \frac{\langle \phi^L\phi^R\rangle}{\Gamma(2\Delta_\phi)} \, \frac{\Theta(-p_+) }{-p_+} \,\left(-4p_+\, \frac{ \cosh \frac{t_3+t_2}{2}}{e^{\frac{t_3 - t_2}{2}}}\right)^{2\Delta_\phi}  \, \exp \left[  -4 i p_+ \, \frac{ \cosh \frac{t_3+t_2}{2}}{e^{\frac{t_3 - t_2}{2}}} -i \pi \Delta_\phi \right] 
\\
\langle \psi^L | p_- \rangle \langle p_- | \psi^L \rangle  
&= \frac{\langle \psi^L \psi^L\rangle}{\Gamma(2\Delta_\psi)} \, \frac{\Theta(-p_-)}{-p_-} \, (-4p_- \, e^{t_1}\sin \delta)^{2\Delta_\psi}\, \exp \left[ 4p_- \,e^{t_1}\sin\delta\right] 
\end{split}
\end{equation}
where the normalization factors are just two-point functions:
\begin{equation}
 \langle \phi^L\phi^R\rangle = \left[ \frac{1}{2\cosh \frac{t_3 + t_2}{2} } \right]^{2\Delta_\phi} \,,\qquad 
 \langle \psi^L \psi^L\rangle = \left[ \frac{1}{2 \sin \delta} \right]^{2\Delta_\psi} \,.
\end{equation}
In order to exponentiate the traversability operator, we treat the scattering events with many copies of the left-right coupling ${\cal O}^L_j {\cal O}_j^R$ as separate according to \eqref{eq:calCdef}, and find:
\begin{equation}
\begin{split}
   \tilde{\cal C} &=   \int dp_+dp_- \, \langle \phi^L | p_+ \rangle \langle p_+ | \phi^R \rangle\langle \psi^L | p_- \rangle \langle p_- | \psi^L \rangle  \; e^{iG p_+ p_-+ig \, \left[ 2\cosh t  - \frac{G}{2} (p_++p_-)  + \frac{ G^2}{4}\, p_+ p_-  \, e^{-t}\right]^{-2\Delta_j}}  \\
   &= \frac{\langle \phi^L\phi^R\rangle\langle \psi^L \psi^L\rangle}{ \Gamma(2\Delta_\phi) \Gamma(2\Delta_\psi)} \int_{-\infty}^0 \frac{dp_+}{-p_+}\int_{-\infty}^0 \frac{dp_-}{-p_-} \, \left(4ip_+\, \frac{ \cosh \frac{t_3+t_2}{2}}{e^{\frac{t_3 - t_2}{2}}}\right)^{2\Delta_\phi}  \, \exp \left[ -4 i p_+ \, \frac{ \cosh \frac{t_3+t_2}{2}}{e^{\frac{t_3 - t_2}{2}}}  \right]   \\
   &\quad \times  \left(-4p_- \, e^{t_1}\sin \delta\right)^{2\Delta_\psi}\, e^{ 4p_- \,e^{t_1}\sin\delta} \; e^{iG p_+ p_-+ig \, \left[ 2\cosh t  - \frac{G}{2} (p_++p_-)  + \frac{ G^2}{4}\, p_+ p_-  \, e^{-t} \right]^{-2\Delta_j}}
  \end{split}
\end{equation}
We change variables to
\begin{equation}
  \tilde{p}_+ = 4p_+\, \frac{ \cosh \frac{t_3+t_2}{2}}{e^{\frac{t_3 - t_2}{2}}} \,,\qquad\qquad \tilde{p}_- = 4p_- \, e^{t_1} \sin \delta\,,
\end{equation}
which gives: 
\begin{equation}
\label{eq:tildeCint}
\begin{split}
   \tilde{\cal C} &= 
   \frac{1}{ \Gamma(2\Delta_\phi) \Gamma(2\Delta_\psi)} \,\langle \phi^L\phi^R\rangle\langle \psi^L \psi^L\rangle \int_{-\infty}^0 \frac{d\tilde{p}_+}{-\tilde{p}_+} (i\tilde{p}_+)^{2\Delta_\phi} \, e^{-i\tilde{p}_+}  \int_{-\infty}^0  \frac{d\tilde{p}_-}{-\tilde{p}_-} \, (-\tilde{p}_-)^{2\Delta_\psi} \, e^{\tilde{p}_-} \\
   &\quad \times \exp \Bigg\{  \frac{iG\,\tilde{p}_+ \tilde{p}_-}{16 \, e^{\frac{t_2-t_3}{2}} \, \cosh \frac{t_3+t_2}{2} \; e^{t_1}\, \sin \delta } \\
   &\qquad \quad\;\;  +ig \, \left[ 2\cosh t  - \frac{G}{8} \left( \frac{\tilde{p}_+\, e^{\frac{t_3-t_2}{2}}}{ \cosh \frac{t_3+t_2}{2}} +\frac{\tilde{p}_-\,e^{-t_1}}{ \sin \delta } \right) + \frac{G^2\, \tilde{p}_+ \tilde{p}_- \, e^{\frac{t_3-t_2}{2}} \,e^{-t_1}}{64 \cosh \frac{t_3+t_2}{2} \sin \delta} \, e^{-t} \right]^{-2\Delta_j} \Bigg\}
  \end{split}
\end{equation}

Let us now assume $\Delta_\psi$ is large so that we can perform the integral over $p_-$ by saddle point.\footnote{ Specifically, we find (for $f(\tilde{p})$ decaying sufficiently quickly):
\begin{equation}
 \frac{1}{\Gamma(2\Delta_\psi)} \int_{-\infty}^0 d\tilde{p}_- \; (-\tilde{p}_-)^{2\Delta_\psi-1} \, e^{\tilde{p}_-} \, f(\tilde{p}_-) \approx f(-2\Delta_\psi) \,.
\end{equation}
} The integral is dominated around $\tilde{p}_- \approx -2\Delta_\psi$ and we get:
\begin{equation}
\begin{split}
   \tilde{\cal C} &= 
   \frac{1}{ \Gamma(2\Delta_\phi)} \,\langle \phi^L\phi^R\rangle\langle \psi^L \psi^L\rangle \int_{-\infty}^0 \frac{d\tilde{p}_+}{-p_+} (i\tilde{p}_+)^{2\Delta_\phi} \,e^{-i(1+ G\Delta_\psi\tilde{z} \, )\tilde{p}_+ }\\
  &\qquad \times \exp \left\{ ig \, \left[ 2  - \frac{G}{4} \left( \frac{\tilde{p}_+\, e^{\frac{t_3-t_2}{2}}}{2  \cosh \frac{t_3+t_2}{2}} -\frac{\Delta_\psi\,e^{-t_1}}{\sin \delta } \right) -\frac{G^2\Delta_\psi}{4}\,  \tilde{z} \,   \,\tilde{p}_+\right]^{-2\Delta_j} \right\}
  \end{split}
\end{equation}
where we set $t=0$ for simplicity and defined
\begin{equation}
 \tilde{z} \equiv \frac{e^{-t_1} \, e^{\frac{t_3-t_2}{2}}}{8 \sin\delta\,\cosh \frac{t_3 + t_2}{2}} \,.
\end{equation}
This expression immediately leads to \eqref{eq:traversableRes}.
We have checked numerically that the saddle point approximation made above is not crucial. Even for small values of $\Delta_\psi$, the above approximation is close to the result obtained by performing both momentum integrals in \eqref{eq:tildeCint} numerically. 

While it is most convenient to perform at least the final integration over $\tilde{p}_+$ numerically, we note that it can in fact be done explicitly in terms of a sum over confluent hypergeometric functions:
\begin{equation}
\label{eq:TraversRes}
\boxed{\;\;
\begin{split}
{\cal C} =  e^{-ig \langle V \rangle_\psi}\,\frac{\langle \phi^L \phi^R \rangle_\psi}{ \langle \phi^L\phi^R \rangle} \; \sum_{n\geq 0}  \frac{(ig\, \langle V \rangle_\psi )^n}{n!} \,\Big\{ Z^{2\Delta_\phi}\, U \left( 2\Delta_\phi, \, 1+2\Delta_\phi-2n\Delta_j, \,Z\right) \Big\}
\end{split}
\;\; }
\end{equation}
with 
\begin{equation}
    Z \equiv -\frac{8i}{G}\frac{\cosh \frac{t_3 + t_2}{2}}{e^{\frac{t_3-t_2}{2}}}\,\frac{2+ \frac{G \Delta_\psi}{4\sin\delta}\, e^{-t_1}}{ 1+ \frac{G\Delta_\psi}{4\sin\delta}\, e^{-t_1}} (1+G\Delta_\psi\tilde{z}) \,.
\end{equation}
Here, we returned to ${\cal C}$ instead of $\tilde{C}$, thus including the normalization factors from \eqref{eq:CCtilde}. Explicitly, the four-point functions appearing above are
\begin{equation}
\begin{split}
 \frac{\langle V \rangle_\psi}{\langle V \rangle} &\equiv 2^{2\Delta_j} \left[ \frac{1}{2+ \frac{G \Delta_\psi}{4\sin\delta} \, e^{-t_1}} \right]^{2\Delta_j} \,, \qquad \quad \frac{\langle \phi^L \phi^R \rangle_\psi}{ \langle \phi^L\phi^R \rangle}   \equiv \left[ \frac{1}{1+G\Delta_\psi\tilde{z}} \right]^{2\Delta_\phi} 
 \,.
 \end{split}
\end{equation}
Note that the result \eqref{eq:TraversRes} has a simple interpretation: the $n$-th term in the sum corresponds to $n$ scattering events with particles created by the ${\cal O}_j^L{\cal O}_j^R$ operator pair. Due to our assumption of the independence of these events, they formally look like a single scattering event involving a particle associated with operator dimension $n\Delta_j$. Let us also remark that the above series expansion in $g$ is not convergent. It formally needs to be resummed before it can be compared to the numerical results of section \ref{sec:numerical}.\footnote{ This is similar to the situation for Lorentzian four-point conformal blocks in two-dimensional CFTs \cite{Chen:2016cms}.}

\paragraph{Factorization of $\tilde{\cal C}_1$:} It is interesting to study the factorization property \eqref{eq:F6factorize} for the traversable wormhole setup. For simplicity, consider the six-point function $\tilde{\cal C}_1$ describing a single scattering event. It can be evaluated using the same methods:
\begin{equation}
\label{eq:C1factor}
\begin{split}
\tilde{\cal C}_1 &= ig\, \langle \psi^L \psi^L \rangle\times \langle V \rangle_\psi\,   \langle \phi^L \phi^R \rangle_\psi\times \left\{ Z^{2\Delta_\phi}\,  U \left( 2\Delta_\phi, \, 1+2\Delta_{\phi j}, \,Z\right) \right\} \,,
\end{split}
\end{equation}
We again observe a natural factorization of $\tilde{\cal C}_1$ into two-point, four-point, and  six-point pieces. Similar to the behavior of ${\cal F}_6(t_1,t_2,0,0)$ in section \ref{sec:twoPert}, the six-point factor (curly bracket in \eqref{eq:C1factor}) interpolates between $1$ and a finite floor value. Note, however, that we only factored out two of the three out-of-time-order four-point functions in \eqref{eq:C1factor}. In the present example, one may wonder if it is natural to factor out the last four-point factor, which now also takes a nontrivial form:
\begin{equation}
    \frac{\langle V \phi^L \phi^R \rangle}{\langle V  \rangle \langle \phi^L \phi^R \rangle} = \left[ -\frac{16i}{G} \,\frac{\cosh \frac{t_3 + t_2}{2}}{e^{\frac{t_3-t_2}{2}}} \right]^{2\Delta_\phi} \;  U \left( 2\Delta_\phi, \, 1+2\Delta_{\phi j}, \,-\frac{16i}{G} \,\frac{\cosh \frac{t_3 + t_2}{2}}{e^{\frac{t_3-t_2}{2}}}\right)\,.
\end{equation}
Factoring out this contribution certainly seems natural from a diagrammatic point of view, as the remaining fully connected six-point factor $\tilde{\cal C}_{1,\text{conn.}}$ will only involve Schwarzian mode exchange diagrams involving all three operators. On the other hand, the resulting `connected' factor will diverge as $t_3-t_2 \gg t_*$. We leave a more detailed investigation of this idea for the future.

\section{Geodesic computation in JT gravity}
\label{app:JTderivation}

In this appendix we explain the geodesic approximation for two-sided boundary correlation functions in JT gravity. In particular we derive the result \eqref{collision_diagnosis} and a similar result for $a=b=0$, which takes the same form as \eqref{eq:F6calc2}.

Let us work in $AdS_2$ embedding space again (see appendix \ref{app:eikonal1} for conventions). The geodesic distance between two boundary points $X$ and $X'$ is given by $d = \log ( -  X \cdot X' )$, where the dot product denotes the flat embedding space inner product with signature $(2,1)$. Our aim is to compute
\begin{align}
\label{eq:F6geodesic}
 {\cal F}_6^\text{(geodesic)}(t_1,t_2;a,b) \equiv e^{-\Delta_j (d_\text{pert.} - d_\text{unpert.})} =\ & \frac{\qty(-\tilde X_R(b)\cdot \tilde X_L(a))^{-\Delta_j}}{\qty(-X_R(b)\cdot  X_L(a))^{-\Delta_j}}\,.
\end{align}
Here, $X_{R,L}$ denote the boundary trajectory without perturbations:
\begin{equation}
\label{eq:XtildeRes}
\begin{split}
	 X_R(t_R) = \qty(1,\, \sinh(t_R), \,\cosh(t_R))\,, \qquad
	X_L(t_L) = \qty(1, \,\sinh(t_L), \,-\cosh(t_L)) \,.
\end{split}
\end{equation}
Similarly, $\tilde{X}_{R,L}$ denote the boundary trajectories after the perturbations with $W_{1,2}$. In order to solve for the perturbed boundary trajectory, we treat the boundary as a charged particle moving in a electric field \cite{Maldacena:2017axo}. The equation of motion satisfied by the right boundary particle is given by
\begin{align}
\label{XEom}
	C\ddot X_R-2E_R X_R = Q_R \,,
\end{align}
where $Q_R$ is the charge vector of the particle, and $E_R = -\frac{1}{2C} \, Q_R^2$. After the perturbation $W_2(t_2)$ the perturbed charges are: 
\begin{align}
&q_2 = \frac{\delta S_2}{2\pi}\qty(1, \sinh(\frac{2\pi}{\beta}t_2),\cosh(\frac{2\pi}{\beta}t_2))\nonumber\\
	&\tilde Q_R = Q_R-q_2 = -\qty(\frac{2\pi C}{\beta},0,0)-\frac{\delta S_2}{2\pi}\qty(1, \sinh(\frac{2\pi}{\beta}t_2),\cosh(\frac{2\pi}{\beta}t_2))\nonumber\\
	&\tilde E_R =-\frac{\tilde Q_R^2}{2C} = \frac{2\pi^2 C}{\beta^2}+\frac{\delta S_2}{\beta}
\end{align}
where $q_2$ is the matter charge due to the perturbations $W_2$. 
Solving \eqref{XEom}, we get
{\small
\begin{align}
\label{trajectory_right}
	\tilde X_R(t_R) = -\frac{\tilde Q_R}{2\tilde E_R}+\qty(X_R(t_2)+\frac{\tilde Q_R}{2\tilde E_R})\cosh(\sqrt{\frac{2\tilde E_R}{C}}(t_R-t_2))+\sqrt{\frac{C}{2\tilde E_R}}\dot X_R(t_2)\sinh(\sqrt{\frac{2\tilde E_R}{C}}(t_R-t_2))
\end{align}
}\normalsize
Similarly, the left boundary trajectory after the perturbation $W_1(t_1)$ is given by
{\small
\begin{align}
\label{trajectory_left}
	\tilde X_L(t_L) = \frac{\tilde Q_L}{2\tilde E_L}+\qty(X_L(t_1)-\frac{\tilde Q_L}{2\tilde E_L})\cosh(\sqrt{\frac{2\tilde E_L}{C}}(t_L-t_1))+\sqrt{\frac{C}{2\tilde E_L}}\dot X_L(t_1)\sinh(\sqrt{\frac{2\tilde E_L}{C}}(t_L-t_1))
\end{align}
}\normalsize
where 
\begin{align}
&q_1 = -\frac{\delta S_1}{2\pi}\qty(1, \sinh(\frac{2\pi}{\beta}t_1),-\cosh(\frac{2\pi}{\beta}t_1)) \nonumber\\
	&\tilde Q_L = Q_L-q_1 = \qty(\frac{2\pi C}{\beta},0,0)+\frac{\delta S_1}{2\pi}\qty(1, \sinh(\frac{2\pi}{\beta}t_1),-\cosh(\frac{2\pi}{\beta}t_1))\nonumber\\
	&\tilde E_R =-\frac{\tilde Q_L^2}{2C} = \frac{2\pi^2 C}{\beta^2}+\frac{\delta S_1}{\beta}
\end{align}
Plugging \eqref{trajectory_right} and \eqref{trajectory_left} into \eqref{eq:F6geodesic}, we get 
\begin{equation}
\label{eq:JTfinalResult}
\begin{split}
{\cal F}_6^\text{(geodesic)} 
   &=
\left[ \frac{1}{1+\frac{\delta S_1}{2(S-S_0)} \, z'_1} \right]^{2\Delta_j} \left[ \frac{1}{1+ \frac{\delta S_2}{2(S-S_0)} \, z'_2} \right]^{2\Delta_j} 
\left[ \frac{1}{1+ \frac{ \frac{\delta S_1 \, \delta S_2}{4(S-S_0)^2} \, z'_1 \, z'_2 \, e^{a+b}}{ \left( 1+ \frac{\delta S_1}{2(S-S_0)} \, z'_1 \right) \left( 1+ \frac{\delta S_2}{2(S-S_0)} \, z'_2 \right)} }
\right]^{2\Delta_j} 
\end{split}
\end{equation}
where
\begin{equation}
\label{eq:z12def}
  z'_1 = \frac{e^{\frac{a-b}{2}-t_1} }{2\, \cosh \frac{a+b}{2}} \,,\qquad
  z'_2 = \frac{e^{-\frac{a-b}{2}-t_2} }{2\, \cosh \frac{a+b}{2}}  \,.
\end{equation}
We ignored terms which are proportional to $\frac{\delta S_i}{S-S_0}$ but are not enhanced by $e^{-t_i}$. 
This result has the same form as the eikonal saddle point result \eqref{eq:F6calc2appendix}.
From the above expressions one immediately obtains \eqref{collision_diagnosis} in the limit where $a,b \gg -t_{1,2}$. Similarly, when $a=b=0$, \eqref{eq:JTfinalResult} yields precisely the same as the saddle point expression \eqref{eq:F6calc2} with operator dimensions replaced by entropy differentials according to \eqref{eq:SDelta}.

\section{Geodesic computation in $AdS_3$ gravity}
\label{app:3D geometry}

Here we explain how to obtain the $\text{AdS}_3$ approximation to the six-point function. In the probe limit, two of the operators simply measure the geodesic distance between points on opposing conformal boundaries. They do so in a geometry sourced by the rest of the operators. In particular, on a BTZ background (or $\text{AdS}_3$-Rindler, as our approximation will not distinguish the two), the operators $W_{1}(t_{1})$ and $W_{2}(t_{2})$ locally insert energy at $t_{1}$ on the left and $t_{2}$ on the right respectively (see figure \ref{fig:bulkSetup}). We approximate the resultant geometry as being that of two null shocks on the background spacetime.

We expect that the increase in distance caused by two sources is approximately the sum of the increases in distance caused by solely the individual sources and some non-linear correction. This is equivalent to saying that a natural structure for the decorrelation of the six-point function is given by product of the four-point functions and some \emph{multiplicative} correction, a non-trivial statement from the boundary perspective \cite{Anous:2020vtw}.

\subsection{Colliding shocks in $AdS_3$}

The corresponding full solution to Einstein's equations pastes together geometries of different masses along the shocks, with the caveat that the mass of the post-collision region is fixed by the other regions so as to avoid a conical singularity. The different symmetric geometries are specified by their radius $R$ and in three dimensions take the form (in AdS units)
\begin{equation}
ds^{2} =-\frac{4}{\left(1+UV\right)^{2}}\, dUdV +r^2 d\phi^{2} ,\quad  r=R\, \frac{1-UV}{1+UV} ,\quad R^{2} = 8\pi G M \,,
\end{equation}
where the geometry under the identification $\phi \sim \phi + 2\pi$ corresponds to BTZ and AdS-Rindler otherwise. Points on the conformal boundary satisfy $UV=-1$, thus all points on the left may be written in terms of their positive null coordinate $U_{L}=e^{R t_{L}}$ and similarly for the right $V_{R} = e^{R t_{R}}$.

A shock fired from the left at $U = e^{R t_{1}}$ collides with a shock fired from the right at $V = e^{R t_{2}}$ at the radius
\begin{equation}
\label{eq:collision radius}
r_{c} = R\,\tanh(R\frac{\abs{t_{1}+t_{2}}}{2})
\end{equation}
and thereby separates out four regions:
\begin{enumerate}
\item Background region $(M,R)$
\item ${W_{1}}(t_{1})$ shock exterior left region $ ( M_{1} , R_{1} )$
$$ M_{1} = M + E_{1} \geq M , \; E_{1} \sim m_{1}$$
\item ${W_{2}}(t_{2})$ shock exterior right region $ ( M_{2} , R_{2} )$ 
$$ M_{2} = M + E_{2} \geq M, \; E_{2} \sim m_{2}$$
\item Post-collision region $ (M_{PC},R_{PC}) $ within the future interior 
$$M_{PC} \geq M + E_{1} + E_{2}$$
\end{enumerate}
The different regions are pasted together so that the radii of the transverse circle is continuous across the shock. Next, a relative boost ambiguity is fixed by requiring continuity of the asymptotic coordinates.  Finally, in order for the geometry be smooth at the collision, a series of boosts between the regions must return to the same frame \cite{Langlois_2002}. Consequently, the mass of our post-collision region gets inflated to much larger values based on how close the collision occurs to the horizon:
\begin{equation}
\label{eq:post-collision mass}
M_{PC} =(M + E_{1} + E_{2}) + \frac{E_{1} E_{2}}{M} \cosh(R \frac{t_{1}+t_{2}}{2})^{2} 	\,.
\end{equation}
Note that the first term is the background mass and total energy added to the system, while the new term is additional mass that becomes significant at the scrambling time of the two shocks. This produces a solution to Einstein's equations for a geometry with a stress tensor localized to the location of the shocks.

\subsection{Geodesics in two-shock $AdS_{3}$}
Our strategy to calculate the geodesic going between $t_L=a$ and $t_R=b$ in such a piecewise-continuous geometry will follow \cite{Roberts:2014isa}. The spatial geodesic will either pass through the background region or post-collision region but not both. We begin by solving for the geodesic when it passes through the background region. We will find that the final geodesic distance determined this way even applies when the actual geodesic goes through the other region instead! In this regime, various local quantities one solves for will become unphysical (e.g. positive distances become negative), but the global geodesic distance will remain as the same physical function.

We begin in the left $W_1(t_1)$ shock exterior region with the $t_L=a$ boundary point $(U_L = e^{R_1 a},V_L=-1/U_L)$ and geodesically connect it through the left exterior region to an arbitrary point on the $W_1(t_1)$ shock: $(U_L = e^{R_1 t_{1}}, r = r_1)$. We do the same with the other side, geodesically connecting the $t_R=b$ boundary point $(U_R = -1/V_R,V_R=e^{R_2 b})$ through the right exterior region to an arbitrary point on the $W_2(t_2)$ shock: $(V_R = e^{R_2 t_{2}}, r = r_2)$. These two arbitrary points specified by $r_1,r_2$ are also in the background region at $(U=e^{R t_{1}}, r=r_1)$ and $(V=e^{R t_{2}}, r=r_2)$, respectively. We complete a path between $t_L=a$ and $t_R=b$ by geodesically connecting the two arbitrary $r_1,r_2$ points through the background region. We have thus connected three geodesics passing through three regions as well as arbitrary points on each shock. This is a geodesic almost everywhere, except for the two arbitrary points. We then solve for the geodesic between $t_L=a$ and $t_R=b$ by extremizing along $r_1,r_2$ the total distance of these three segments,
\begin{equation}
d_{a,b}^{(W_1W_2)} = \underset{r_1,r_2} {\text{ext}}(d_{a,1} +  d_{2,b} + d_{1,2})\,.
\end{equation}

When regulated, the three-geodesic path passing through the arbitrary points $r_1$ and $r_2$ is comprised of distances
\begin{equation}
\begin{split}
d_{a,1} &= \ln(2 R_1 \frac{\rho_1-r_1}{\rho_1^2-R_1^2}), \quad \rho_1 \equiv R_1 \coth(R_1 \frac{a-t_1}{2}) \\
d_{2,b} &= \ln(2 R_2 \frac{\rho_2-r_2}{\rho_2^2-R_2^2}), \quad \rho_2 \equiv R_2 \coth(R_2 \frac{b-t_2}{2}) \\
d_{1,2} &= \cosh^{-1} \left( 1 + 2 \frac{(r_1-r_c)(r_2-r_c)}{R^2 - r_c^2} \right), \qquad r_c \equiv R \tanh(-R \frac{t_1+t_2}{2}). \\
\end{split}
\end{equation}
Extremizing the distances over $r_1,r_2$ nets the relations
\begin{equation}
\tanh \frac{d_{1,2}}{2} = \frac{r_1-r_c}{\rho_1-r_1} = \frac{r_2-r_c}{\rho_2-r_2}
\end{equation}
This allows us to determine the radii at which the geodesic crosses the shocks in terms of the boundary parameters, bundled together into the functions $r_c(t_1+t_2), \rho_1(a-t_1),\rho_2(b-t_2)$
\begin{equation}
\frac{r^{geo}_1 - r_c}{\rho_1-r_c} = \frac{r^{geo}_2 - r_c}{\rho_2-r_c} = \frac{1}{2}\left(1-\frac{R^2-r_c^2}{(\rho_1-r_c)(\rho_2-r_c)}\right),
\end{equation}
as well as the distance between the two shocks that is crossed by the global geodesic,
\begin{equation}
e^{d_{1,2}^{geo}} = \frac{(\rho_1-r_c)(\rho_2-r_c)}{R^2-r_c^2}\,.
\end{equation}

We thus have that the total distance for this geodesic is given by
\begin{equation}
\begin{split}
e^{d_{a,b}^{(W_1W_2)}} &= \frac{2 R_1 (\rho_1-r_1^{geo})}{\rho_1^2-R_1^2} \frac{2 R_2 (\rho_2-r_2^{geo})}{\rho_2^2-R_2^2} \frac{(\rho_1-r_c)(\rho_2-r_c)}{R^2-r_c^2} \\
&= \frac{R_1 R_2 (R^2-r_c^2)}{(\rho_1^2-R_1^2)(\rho_2^2-R_2^2)} \left(1+\frac{(\rho_1-r_c)(\rho_2-r_c)}{R^2-r_c^2}\right)^2 \,.
\end{split}
\end{equation}
A finite $G$ gives a small but finite shift to the $R_1,R_2$ due to the energy inserted by $W_1,W_2$. We shall parametrize such a change as $ R_i \equiv R(1+\varepsilon_i) $. It is natural in this context to express the entropy changes in terms of the small increases $\varepsilon_{1,2}$ of the horizon radii due to the insertion of the operators:
\begin{equation}
    \varepsilon_1 = \frac{\delta S_1}{S} = \frac{R_1 -R}{R} \,,\qquad \varepsilon_2 = \frac{\delta S_2}{S} = \frac{R_2 -R}{R}\,.
\end{equation}
Thus the nonlinear $G$ physics is contained within the variables $\rho_1,\rho_2$, which corrects both its pre-factor and rate of change. The latter effect manifests in terms that grow with times as $e^{G t}$, and thus we will approximate 
\begin{equation}
\begin{split}
\rho_1 \equiv R_1 \coth(R_1 \frac{a-t_1}{2}) & \approx R_1\coth(R\frac{a-t_1}{2}) + {\cal O}(e^{\varepsilon_1(a-t_1)}) \\
\rho_2 \equiv R_2 \coth(R_2 \frac{b-t_2}{2}) & \approx R_2\coth(R\frac{b-t_2}{2}) + {\cal O}(e^{\varepsilon_2(b-t_2)})
\end{split}
\end{equation}

This small approximation allows us to observe a factorization structure of this total distance. We will find that it is almost the change in distance separately induced by only $W_1$ and $W_2$. To being seeing this, let us set $\varepsilon_1=0$ and $\varepsilon_2=0$ to note that in thermal units the distance between the two points in the background is (up to a regularization constant) 
\begin{equation}
e^{d_0} = \cosh(\frac{a+b}{2})^2
\end{equation}
Similarly, we find that the \emph{change} in distance ($\tilde{d} \equiv d - d_0$) induced by a geometry with either a single $W_1$ shock or a single $W_2$ shock is given by
\begin{equation}
e^{\tilde{d}_{W_1}} = \left( 1 + \varepsilon_1 \frac{\cosh(\frac{a-t_1}{2}) \cosh(\frac{b+t_1}{2})}{\cosh(\frac{a+b}{2})} \right)^2 \quad e^{\tilde{d}_{W_2}} = \left( 1 + \varepsilon_2 \frac{\cosh(\frac{b-t_2}{2}) \cosh(\frac{a+t_2}{2})}{\cosh(\frac{a+b}{2})} \right)^2
\end{equation}
Thus, we see that there is a non-linear effect in our overall distance for the geodesic in the geometry with both a $W_1$ shock and a $W_2$ shock:
\begin{equation}
\label{eq:AdS3FullResult}
\begin{split}
\exp({\tilde{d}_{W_1W_2}}) &= \frac{\exp(\tilde{d}_{W_1}+\tilde{d}_{W_2})}{(1+\varepsilon_1)(1+\varepsilon_2) } \left(1+\frac{\varepsilon_1 \varepsilon_2}{4} \frac{\sinh(a-t_1)\sinh(b-t_2)}{\cosh(\frac{a+b}{2})^2  \exp(\frac{\tilde{d}_{W_1}+\tilde{d}_{W_2}}{2}) } \right)^2
\end{split}
\end{equation}
The increase in distance is almost but also more than just the linear sum of the individual effects of only $W_{1,2}$. Let us take a moment to emphasize that regardless of whether or not the geodesic goes through the post-collision region, the sign of the non-linear correction is given by $ \text{sgn}((a-t_1)(b-t_2)) $. Thus for our configurations, this nonlinear correction is always \emph{positive}.

\section{Comparison of large-$C$ perturbation theory and eikonal method}
\label{sec:perturbative}

In this appendix we unearth how the eikonal resummation of exponentially boosted contributions to the six-point function emerges from a perturbative treatment of the theory of boundary reparametrizations. While we focus on the case of the Schwarzian action, we expect qualitatively similar arguments to apply in higher dimensions.

The `heavy-heavy-light' eikonal result for the six-point function, \eqref{eq:F6calc2}, can also be understood perturbatively in terms of Schwarzian reparametrization mode exchange diagrams. To be slightly more general, let us consider the analogous saddle point result ($\Delta_{1,2}\gg\Delta_j$) for general values of $a,b$, which we reproduce here (see \eqref{eq:F6calc2appendix}):
\begin{equation}
\label{eq:F6calc2gen}
{\cal F}_6 (t_1,t_2;a,b)
   \approx
\left[ \frac{1}{1+\frac{2\Delta_1}{C} \, z_1} \right]^{2\Delta_j} \left[ \frac{1}{1+ \frac{2\Delta_2}{C} \, z_2} \right]^{2\Delta_j} 
\left[ 1+ \frac{ \frac{4\Delta_1\Delta_2}{C^2} \, z_1 \, z_2 \, e^{a+b}}{ \left( 1+ \frac{2\Delta_1}{C} \, z_1 \right) \left( 1+ \frac{2\Delta_2}{C} \, z_2 \right)} 
\right]^{2\Delta_j} \,,
\end{equation}
with $z_{1,2}$ as in \eqref{eq:z12def}, and $C = \frac{2}{G}$.
To investigate this perturbatively, let us first perform an expansion of \eqref{eq:F6calc2gen} in the small parameter $\frac{1}{C}$:
\begin{equation}
\label{eq:F6perturbative}
\begin{split}
  {\cal F}_6 &\approx 1 - \frac{1}{C} \, 4\Delta_j \left( \Delta_1z_1 + \Delta_2 z_2 \right) \\
  & \qquad +\frac{4}{C^2} \Delta_j\left[ \left(\Delta_1^2 z_1^2 +\Delta_2^2 z_2^2\right)(2\Delta_j+1) + \Delta_1 \Delta_2\, (4 \Delta_j - 2 e^{a+b}) z_1 z_2 \right]+ {\cal O}\left( \frac{1}{C^3} \right) 
\end{split}
\end{equation}
Note that this can also be thought of as a triple expansion in the parameters $\{ \frac{\Delta_1}{C}, \, \frac{\Delta_2}{C}, \, \Delta_j\}$. We would like to associate the various terms in this expansion with Schwarzian mode exchange diagrams. The general structure will contain two types of diagrams: `disconnected' diagrams give rise to terms that originate from the four-point function factors in \eqref{eq:F6calc2gen}. On the other hand, `connected' diagrams account for the third factor in \eqref{eq:F6calc2gen}, which is the new six-point piece. One can immediately see from \eqref{eq:F6calc2gen} that connected pieces will always contain at least one power of each of the three operator dimensions.

\subsection*{Reparametrization mode rules}

The perturbative approach has the same starting point as the eikonal computation, i.e., the Schwarzian integral \eqref{eq:F6SchwGen}. The difference is that we now expand the reparametrization $\hat{t}[\hat{u}]$ perturbatively around the thermal saddle. For clarity, let us first work in Euclidean signature and in the end perform the analytic continuation (we denote Euclidean time as $u$ and Lorentzian time as $\hat{u}$). The reparametrization of the saddle then takes the form ${t}_\epsilon[{u}] = \tan \left[ \frac{1}{2} \left({u} + \epsilon + \frac{1}{2} \epsilon \epsilon' + \ldots \right) \right]$, where $\epsilon({u})$ is the small reparametrization. The bilocal operators \eqref{eq:Bdef} can be expanded in $\epsilon$ accordingly. In the present context we find it more conventional to work with normalized bilocal operators, whose expansion takes the following form \cite{Anous:2020vtw}:
\begin{equation}
\label{eq:Bexpand}
\begin{split}
 {\cal B}_{\Delta}({u}_a,{u}_b) \equiv \frac{G_\Delta(u_a,u_b)}{\langle {\cal O}_\Delta(u_a) {\cal O}_\Delta(u_b)\rangle} &=  \exp \left[ \sum_{q \geq 1} \, b^{(q)}_\Delta(u_a,u_b) \right] = 1 + b_\Delta^{(1)} + \left[ \frac{1}{2!} \, \left( b_\Delta^{(1)} \right)^2 + b^{(2)} \right] + \ldots
\end{split}
\end{equation}
where the exponential structure comes about when one attempts to expand in `atomic' building blocks $b^{(q)}_\Delta$, all of which are proportional to a single power of $\Delta$:
\begin{equation}
\label{eq:bqDef}
\begin{split}
 b^{(1)}_\Delta(u_a,u_b) &\equiv\; \includegraphics[width=.08\textwidth]{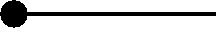} \; =  \Delta \left[ \epsilon_a' + \epsilon_b' - \frac{\epsilon_a - \epsilon_b}{\tan \frac{u_{ab}}{2}} \right]  \,,\\
 b^{(2)}_\Delta(u_a,u_b) &\equiv\;
 \begin{gathered}
 \includegraphics[width=.08\textwidth]{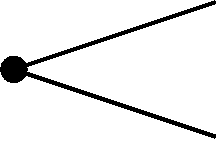}
 \end{gathered} 
 \;= \frac{\Delta}{2} \left[  \epsilon_a''\epsilon_a + \epsilon_b''\epsilon_b - \frac{\epsilon_a'\epsilon_a - \epsilon_b'\epsilon_b}{\tan \frac{u_{ab}}{2}} + \frac{(\epsilon_a - \epsilon_b)^2}{2\sin^2 \left( \frac{u_{ab}}{2} \right) } \right] \,, \;\;\ldots
\end{split}
\end{equation}
where $\epsilon_a \equiv \epsilon(u_a)$ etc., and we use a simple diagrammatic notation from \cite{Anous:2019yku}. In these diagrams, each `node' corresponds to a power of $\Delta$ associated with the bilocal operator insertion, and each line corresponds to a reparametrization mode $\epsilon$.

The propagator and interaction vertices of the Schwarzian mode follow from the expansion of the Schwarzian action. In Euclidean signature and up to total derivatives:
\begin{equation}
\label{eq:SchwExpand}
  -S[t(u)]  = C \int_0^\beta du \; \left[  \text{const.} - \frac{1}{2} \epsilon\, (\epsilon'' + \epsilon'''') + \frac{1}{8} \epsilon' \epsilon'( 2 \epsilon'  +\epsilon''' ) + \ldots \right]
\end{equation}
The propagator is well known \cite{Maldacena:2016upp}:
\begin{equation}
\label{eq:epsProp}
  \langle \epsilon(u) \epsilon(0) \rangle = \frac{1}{2\pi C} \sum_{|n| \geq 2} \frac{e^{inu}}{n^2(n^2-1)} = \frac{1}{2\pi C} \left[ \frac{2\sin u-(\pi+u)}{2}\, (\pi+u) +2\pi \, \Theta(u) \left(u-\sin u\right)\right] \,.
\end{equation}
The three-point vertex follows from \eqref{eq:SchwExpand} in a similar manner \cite{Qi:2019gny}:
\begin{equation}
\label{eq:epsVert}
  \langle \epsilon(u_a) \epsilon(u_b) \epsilon(u_c) \rangle = \frac{i}{8 \pi^2C^2}\sum_{\substack{|m|, |n| \geq 2 \\ |m+n| \geq 2}} \frac{mn(m+n)(m^2+mn+n^2-3)\, e^{i(n u_{ac} + m u_{bc})}}{n^2(n^2-1)m^2(m^2-1)(m+n)^2((m+n)^2+1)} 
\end{equation}
This can be evaluated straightforwardly, but is not very pleasant to look at in position space.

\subsection*{Six-point function to second order}

The six-point function can now be expanded in terms of bilocal blocks $b^{(q)}_\Delta$. Even just expanding $ {\cal F}_6 = {\langle}  {\cal B}_{\Delta_1}(\hat{u}_1,\hat{u}_2) \, {\cal B}_{\Delta_j}(\hat{u}_3,\hat{u}_4) \, {\cal B}_{\Delta_2} (\hat{u}_5,\hat{u}_6) {\rangle} $ to order ${\cal O}(C^{-2})$ gives a rather large number of terms. However, these can be repackaged in a compact way by noting that most of the terms are `disconnected' in the sense that they only involve Schwarzian mode exchanges between a subset of the bilocals. 
These disconnected contributions are precisely of the form that one expects from expectation values of two or one bilocals. More precisely:
\begin{equation}
\label{F6:restructured}
\begin{split}
 {\cal F}_6 
&= {\langle}  {\cal B}_{\Delta_1} \, {\cal B}_{\Delta_j} \, {\cal B}_{\Delta_2}  {\rangle} \\
&= \frac{\big{\langle}  {\cal B}_{\Delta_1}\, {\cal B}_{\Delta_j} \big{\rangle} \big{\langle}  {\cal B}_{\Delta_j}\, {\cal B}_{\Delta_2} \big{\rangle} \big{\langle}  {\cal B}_{\Delta_1}\, {\cal B}_{\Delta_2} \big{\rangle} }{\big{\langle}  {\cal B}_{\Delta_1} \big{\rangle} \big{\langle}  {\cal B}_{\Delta_j} \big{\rangle}\big{\langle}  {\cal B}_{\Delta_2} \big{\rangle}} \\
&\quad \times  \left[1+ \langle b^{(1)}_{\Delta_1} b^{(1)}_{\Delta_j} b^{(1)}_{\Delta_2} \rangle+ \langle b^{(2)}_{\Delta_1} b^{(1)}_{\Delta_j} b^{(1)}_{\Delta_2} \rangle_\text{conn.} +  \langle b^{(1)}_{\Delta_1} b^{(2)}_{\Delta_j} b^{(1)}_{\Delta_2} \rangle_\text{conn.} +  \langle b^{(1)}_{\Delta_1} b^{(1)}_{\Delta_j} b^{(2)}_{\Delta_2} \rangle_\text{conn.} + \ldots \right]
\end{split}
\end{equation}
where we suppress time arguments and ``conn.'' instructs us to extract the piece which connects all three bilocals. For example:
\begin{equation}
\langle b^{(2)}_{\Delta_1} b^{(1)}_{\Delta_j} b^{(1)}_{\Delta_2} \rangle_\text{conn.} \equiv \langle b^{(2)}_{\Delta_1} b^{(1)}_{\Delta_j} b^{(1)}_{\Delta_2} \rangle - \langle b^{(2)}_{\Delta_1} \rangle \langle b^{(1)}_{\Delta_j} b^{(1)}_{\Delta_2} \rangle 
\equiv
 \begin{gathered}
 \includegraphics[width=.1\textwidth]{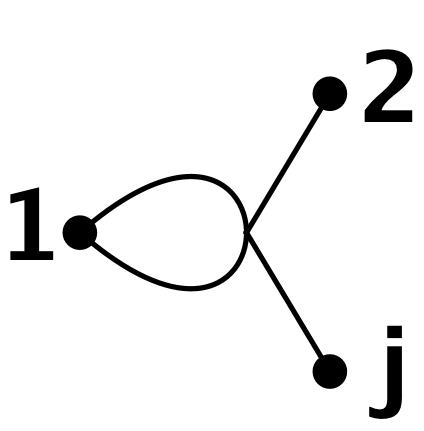}
 \end{gathered} 
 -\;
  \begin{gathered}
 \includegraphics[width=.1\textwidth]{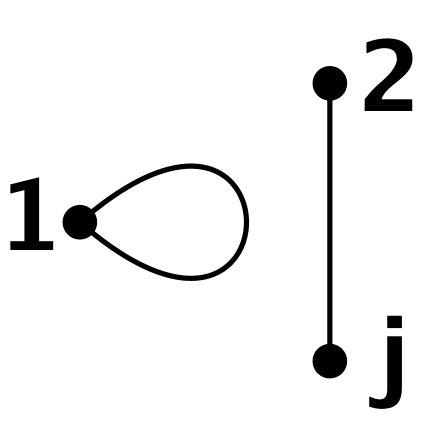}
 \end{gathered} 
 \equiv 
 \begin{gathered}
 \includegraphics[width=.1\textwidth]{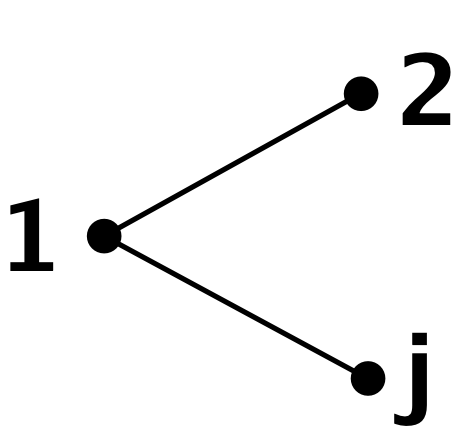}
 \end{gathered} 
 + {\cal O}(C^{-3})
\end{equation}
The first term involves four-point functions $\langle \epsilon\epsilon\epsilon\epsilon\rangle$ of the Schwarzian mode, which need to be Wick contracted into products of two-point functions in order to obtain the leading large-$C$ contribution (we ignore $1/C$ corrections). 
One can easily verify that (at least at the level of abstract diagrammatics) the structure of \eqref{F6:restructured} holds to arbitrary orders: two- and four-point function contributions can be factored out and one is left with the last line, which captures the fully connected, genuinely six-point Schwarzian mode exchanges. In terms of Schwarzian mode exchange diagrams, these connected contributions correspond to diagrams which connect all three pairs of probe operators and cannot be written as products of diagrams connecting only two of the pairs each. And indeed, this is also the structure we found for the exponentially growing contributions in the eikonal calculation. Let us now corroborate this connection in some more detail.

\paragraph{Connected contributions:}
First, let us discuss the most interesting contribution to \eqref{eq:F6calc2gen}, which comes from connected diagrams in the sense described above. The first {\it connected} term at this order is:
\begin{equation} 
\begin{split}
 \langle b^{(1)}_{\Delta_1} b^{(2)}_{\Delta_j} b^{(1)}_{\Delta_2} \rangle_\text{conn.} &= - \frac{\Delta_j \Delta_1 \Delta_2}{2C^2\sin \delta_1 \sin \delta_2 } \, \frac{(a-t_1-\sinh(a-t_1))(b-t_2-\sinh(b-t_2))}{\cosh^2 \frac{a+b}{2}} + {\cal O}(\delta_{1,2}^{-1})
 \end{split}
\end{equation}
which is obtained by extracting the `connected' Wick contractions (where $b^{(2)}_{\Delta_j}$ is contracted with both $b^{(1)}_{\Delta_1}$ and $b^{(1)}_{\Delta_2}$, as opposed to with itself) and evaluating them using the propagator \eqref{eq:epsProp}.
For large $a-t_1 \gg 1$ and $b-t_2\gg 1$, the exponentially growing contribution is just
\begin{equation} 
\begin{split}
 \langle b^{(1)}_{\Delta_1} b^{(2)}_{\Delta_j} b^{(1)}_{\Delta_2} \rangle_\text{conn.} \Big{|}_{\text{exp.}} = - \frac{\Delta_j \Delta_1 \Delta_2}{C^2} \, 8\, e^{a+b}\, z_1 z_2 \,.
 \end{split}
\end{equation}
This reproduces one of the terms in the second line of \eqref{eq:F6perturbative}. 
One can check that the same diagrams with permuted insertion points ($\langle b^{(2)}_{\Delta_1} b^{(1)}_{\Delta_j} b^{(1)}_{\Delta_2} \rangle_\text{conn.} $ and $\langle b^{(1)}_{\Delta_1} b^{(1)}_{\Delta_j} b^{(2)}_{\Delta_2} \rangle_\text{conn.} $) do not give any exponentially growing contributions. 

The last remaining {\it connected} contribution in \eqref{F6:restructured} is of the form
\begin{equation}
   \langle b^{(1)}_{\Delta_1} b^{(1)}_{\Delta_j} b^{(1)}_{\Delta_2} \rangle \equiv\;
   \begin{gathered}
\includegraphics[width=.1\textwidth]{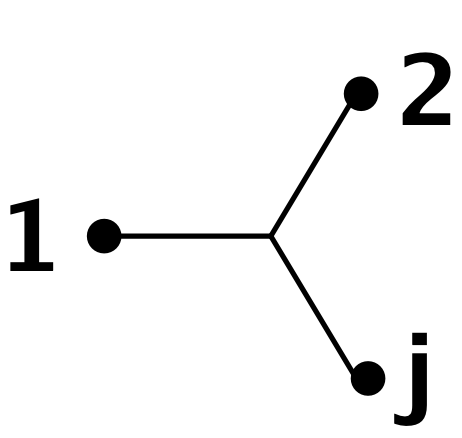} \end{gathered}
\end{equation}
and can be computed using the vertex \eqref{eq:epsVert}. Its general form is complicated, but one can check that it does not exhibit exponential growth (nor does it contribute at leading order in the small-$\delta$ expansion).\footnote{ This is consistent with analogous observations regarding the global six-point identity conformal block in two-dimensional CFTs \cite{Anous:2020vtw}.}

\paragraph{Disconnected contributions:}
The disconnected contributions to \eqref{eq:F6calc2gen} originate from the four- and two-point factors in the second line of \eqref{F6:restructured}. Obviously, the terms at order ${\cal O}(C^{-1})$ are all disconnected; indeed, the first four-point pre-factor of \eqref{eq:F6calc2gen} is easy to compute and the result is well known \cite{Maldacena:2016upp}. Its exponentially growing contribution takes the form
\begin{equation}
\langle b^{(1)}_{\Delta_1} b^{(1)}_{\Delta_j} \rangle \Big{|}_\text{exp.} \equiv \;
   \begin{gathered}
\includegraphics[width=.1\textwidth]{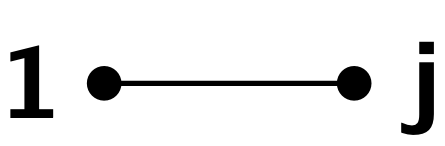} \end{gathered} \;\Big{|}_\text{exp.}
= -\frac{4}{C} \, \Delta_j \Delta_1 z_1 \,,
\end{equation}
which we recognize as one of the terms in the expansion \eqref{eq:F6perturbative}. The term proportional to $\Delta_j \Delta_2$ follows similarly. The third four-point factor, $\langle  {\cal B}_{\Delta_1}\, {\cal B}_{\Delta_2} \rangle$, does not exhibit exponential growth (the operators involved are always in-time-order) and therefore makes no appearance in \eqref{eq:F6calc2gen}.

Next, we consider disconnected contributions at ${\cal O}(C^{-2})$. There are quite a few such terms. We shall only write the ones, which give rise to exponentially growing contributions. In particular, the four-point pre-factors contain terms such as 
\begin{equation}
\begin{split}
 \frac{1}{2!} \langle b^{(1)}_{\Delta_1}b^{(1)}_{\Delta_1} b^{(1)}_{\Delta_j}  \rangle+ \frac{1}{2!} \langle b^{(1)}_{\Delta_1}b^{(1)}_{\Delta_1} b^{(2)}_{\Delta_j}  \rangle \Big{|}_{\text{exp.}} &\sim \frac{4}{C^2} \, \Delta_j\left(  \Delta_1 z_1 \right)^2 + \ldots \,, \\
\frac{1}{(2!)^2} \langle b^{(1)}_{\Delta_1} b^{(1)}_{\Delta_1} b^{(1)}_{\Delta_j}b^{(1)}_{\Delta_j} \rangle  \Big{|}_\text{exp.} \supset 2\times\left( \frac{1}{2!} \,  \langle b^{(1)}_{\Delta_1} b^{(1)}_{\Delta_j} \rangle \right)^2 \Big{|}_\text{exp.} &\sim \frac{8}{C^2} \, \left( \Delta_j \Delta_1 z_1 \right)^2+ \ldots \,, \\
 \frac{1}{2!} \langle b^{(1)}_{\Delta_1}b^{(1)}_{\Delta_j} b^{(1)}_{\Delta_j}b^{(1)}_{\Delta_2}  \rangle \Big{|}_{\text{exp.}} \supset
   \langle b^{(1)}_{\Delta_1}b^{(1)}_{\Delta_j}\rangle \langle b^{(1)}_{\Delta_j}b^{(1)}_{\Delta_2}  \rangle \Big{|}_{\text{exp.}}  &\sim \frac{16}{C^2} \, \Delta_j^2  \Delta_1 \Delta_2 z_1 z_2 + \ldots\,,
\end{split}
\end{equation} 
where the omitted terms (`$...$') do not lead to exponential growth in any of the configurations we consider. We evaluated these expressions just using the propagator \eqref{eq:epsProp} and the vertex \eqref{eq:epsVert}. Note that we count the third term as `disconnected' despite the fact that it involves all three operator dimensions; this is because it is of the form of a product of lower order contributions, each of which involves only two out of the three bilocals. 

Adding up the above terms completes the perturbative computation of \eqref{eq:F6perturbative} to order ${\cal O}(C^{-2})$. There are two interesting physical lessons: $(i)$  the organization of diagrams into connected and disconnected pieces, \eqref{F6:restructured}, is a very intriguing and efficient structure and shows that the connected contributions are relatively few and simple. $(ii)$ exponential decay of the connected piece and due the out-of-time-order arrangement originates from Schwarzian mode `ladder diagrams'; self-interaction vertices are not important. This makes the calculation tractable at higher orders and underlies the simplicity of the eikonal approach.

\bibliographystyle{utphys}
\bibliography{references}

\end{document}